\documentclass{article}

 \usepackage[preprint]{neurips_2026}

\usepackage[utf8]{inputenc}
\usepackage[T1]{fontenc}

\usepackage{xcolor}
\definecolor{DarkGreen}{rgb}{0.1,0.5,0.1}
\definecolor{DarkRed}{rgb}{0.5,0.1,0.1}
\definecolor{DarkBlue}{rgb}{0.1,0.1,0.5}

\usepackage{hyperref}
\hypersetup{
    colorlinks=true,       
    linkcolor=DarkBlue,          
    citecolor=DarkGreen,        
    filecolor=DarkRed,    
    urlcolor=DarkBlue,          
}

\usepackage{xurl}
\usepackage{microtype}
\usepackage{comment}

\usepackage[ruled]{algorithm2e}
\renewcommand{\algorithmcfname}{ALGORITHM}

\usepackage{booktabs, tabularx}

\usepackage{enumitem}
\setlist{leftmargin=.5cm}

\usepackage{subcaption}
\usepackage{amsmath, mathtools, amssymb, nicefrac, bm}
\allowdisplaybreaks

\usepackage{soul}

\usepackage{amsthm, thm-restate}

\newtheorem{theorem}{Theorem}[section]
\newtheorem{definition}{Definition}
\newtheorem{lemma}[theorem]{Lemma}
\newtheorem{assumption}{Assumption}
\newtheorem{strategy}{Parameterization}
\newtheorem{game}{Game}

\newtheorem{algorithmdef}{Algorithm}
\newtheorem{remark}[theorem]{Remark}

\usepackage[nameinlink]{cleveref}
\crefname{tp}{protocol}{protocols}
\Crefname{tp}{Protocol}{Protocols}
\crefname{game}{game}{games}
\Crefname{game}{Game}{Games}
\crefname{strategy}{parameterization}{parameterizations}
\Crefname{strategy}{Parameterization}{Parameterizations}
\crefname{assumption}{assumption}{asssumptions}
\Crefname{assumption}{Assumption}{Assumptions}
\crefname{definition}{def.}{defs.}
\Crefname{definition}{Definition}{Definitions}
\crefname{algorithmdef}{alg.}{algs.}
\Crefname{algorithmdef}{Algorithm}{Algorithms}

\usepackage{enumitem}
\usepackage{nameref}
\newcommand{\gameref}[1]{\hyperref[#1]{game \nameref*{#1}}}
\newcommand{\Gameref}[1]{\hyperref[#1]{Game \nameref*{#1}}}

\usepackage{crossreftools}
\let\phi\varphi%
\let\e\varepsilon%

\newcommand{\lrb}[1]{\left(#1\right)} 
\newcommand{\lsb}[1]{\left[#1\right]} 
\newcommand{\lcb}[1]{\left\{#1\right\}} 
\newcommand{\labs}[1]{\left\lvert#1\right\rvert} 

\newcommand{\E}[1]{\mathbb{E}\!\lsb{#1}}
\newcommand{\Et}[1]{\mathbb{E}_t\!\lsb{#1}}
\newcommand{\Pb}[1]{\mathbb{P}\!\lrb{#1}}

\newcommand{\inv}{I}
\newcommand{\cash}{C}
\newcommand{\wealth}{W}

\newcommand{\strat}{\pi}

\newcommand{\taker}{{\mathbb{T}}}
\newcommand{\ka}{k_\alpha}
\newcommand{\kb}{k_\beta}

\newcommand{\maker}{{\mathbb{M}}}
\newcommand{\va}{v_\alpha}
\newcommand{\vb}{v_\beta}

\newcommand{\noise}{\e}

\renewcommand{\tilde}{\widetilde}

\title{The Invisible Handshake:\\     Persistent Overpricing by Adaptive Market Agents}

\author{%
  Luigi Foscari \\
  Department of Computer Science\\
  University of Milan \\
  \And
  Emanuele Guidotti \\
  Institute of Finance \\
  University of Lugano \& LLUI \\
  \And
  Nicolò Cesa-Bianchi \\
  Department of Computer Science\\
  University of Milan \\
  \And
  Tatjana Chavdarova \\
  Department of Computer Science \\
  Vienna University of Technology \\
  \And
  Alfio Ferrara \\
  Department of Computer Science\\
  University of Milan \\
}

\begin{document}

\maketitle

\begin{abstract}
We study overpricing in a repeated game between two representative agents: a market maker, who controls market liquidity, and a market taker, who chooses trade quantities. 
Market prices evolve through the endogenous price impact of trades and exogenous shocks.
We define overpricing relative to a counterfactual price path that holds fixed the same sequence of shocks while shutting down price impact, and characterize the set of feasible strategy profiles that generate persistent overpricing while respecting cash and inventory constraints.
%
We provide a sufficient condition for decentralized learning to reach the overpricing region in finite time, and we show that this condition is satisfied, in particular, by projected stochastic gradient ascent.
A key step in the analysis is a decomposition of the game into a competitive component, which favors zero price impact, and a collaborative component, which makes overpricing jointly profitable when aggregate inventory is positive.
%
We further show that the same structural incentives govern both myopic and farsighted objectives.
%
Together, these results show how decentralized learning by adaptive market agents can lead to persistent overpricing in financial markets.
\end{abstract}

\section{Introduction}

Markets work well when prices provide accurate signals for resource allocation. For example, if wheat prices rise because poor harvests are expected, farmers may plant more wheat; if prices rise without any change in fundamentals, they may overproduce. 
As AI systems designed to maximize the value of their portfolios are increasingly deployed in financial markets, a central question is whether adaptive market agents can themselves become a source of persistent price distortions.\footnote{Recent policy reports have highlighted several risks associated with the growing use of AI in financial markets. See Chapter III of the 2024 Annual Economic Report of the Bank for International Settlements (\url{https://www.bis.org/publ/arpdf/ar2024e.pdf}) and Chapter 3 of the 2024 Global Financial Stability Report by the International Monetary Fund (\url{https://www.imf.org/-/media/Files/Publications/GFSR/2024/October/English/textrevised.ashx}).} 

This paper studies whether decentralized learning by such agents can generate persistent overpricing relative to the path implied by fundamentals alone.
To do so, we introduce a \emph{repeated stochastic game} between two representative agents: a \emph{market maker} and a \emph{market taker}. The two agents represent the two sides of the market. The maker controls market liquidity by choosing price-impact parameters, while the taker chooses trade quantities. Both agents hold cash and inventory and trade to maximize the value of their own portfolios, subject to constraints that rule out short and leveraged positions. Prices have two components: an exogenous component driven by shocks to fundamentals and an endogenous component generated by the price impact of trades. We measure overpricing by comparing the realized market price path to a counterfactual price path that holds fixed the same sequence of shocks while shutting down price impact.

Within this framework, we show that persistent overpricing can arise endogenously from decentralized learning alone, without any communication or explicit coordination. The key mechanism is structural: when aggregate inventory is positive, higher prices increase aggregate portfolio value, creating a collaborative component in otherwise non-collaborative learning objectives.

\paragraph{Contributions.}
Our contributions are threefold.
\begin{itemize}
    \item We characterize feasible overpricing profiles: we introduce a repeated game with endogenous price impact and cash-inventory constraints, derive a low-dimensional parameterization of strategies, and identify the region of the parameter space that generates persistent overpricing (\cref{sec:model}).
    \item Through a strategic decomposition of the stage game, we show that the players' incentives contain a collaborative component that favors persistent overpricing. We then prove that decentralized learning reaches the overpricing region in finite time for adaptive updates satisfying a stochastic progress condition, and verify this condition for projected stochastic gradient ascent (\cref{sec:learnbabylearn}).
    \item We show that this phenomenon is not purely myopic: agents optimizing infinite-horizon wealth face the same overpricing incentives as greedy myopic agents and reach the same overpricing region of the parameter space (\cref{sec:farsighted}).
\end{itemize}

\paragraph{Modeling remarks.}
Our model is a two-player game where the two players should be interpreted as the two sides of the market rather than as two individual market participants. This modeling choice is in the spirit of \cite{kyle_continuous_1985}: the maker represents aggregate liquidity provision, while the taker represents aggregate order flow. We impose a fair-pricing condition to model competitive liquidity provision in reduced form, as in a market with many competing makers. We rule out short and leveraged positions for the same reason: such positions may exist for individual traders, but they offset in aggregate. Hence the representative maker and taker hold nonnegative aggregate cash and inventory positions. These constraints introduce technical challenges that we briefly discuss below.

\paragraph{Technical challenges.}
A natural formulation for the problem is a constrained stochastic game \citep{shapley1953}: agents' feasible trades depend on their current cash and inventory, prices depend endogenously on both agents' actions and learning updates must remain feasible along the entire realized trajectory. 
Direct Markov-game analysis is therefore difficult.
Our approach is to exploit the structure of the trading problem: we design a strategy parameterization that guarantees admissible market moves, separates exogenous fundamental shocks from endogenous price impact and reduces the learning problem to a tractable constrained dynamical system. 
This reduction is what makes it possible to prove finite-time entry into the overpricing strategy region for a broad class of stochastic adaptive updates, and then to show that the same incentive survives in the infinite-horizon objective.

\paragraph{Related work.}
Our paper connects to work on learning-based market making, including online and reinforcement-learning approaches to quoting and liquidity provision \citep{abernethy2013adaptive,10.5555/3237383.3237450}, as well as work on learning-based market taking, such as online portfolio selection \citep{cover2,cover1,hazan_online_2015}.
More broadly, our setting fits within multi-agent reinforcement learning, where convergence is usually hard to characterize outside special cases \citep{littman94,hu_wellman_1998,albrecht_multi-agent_nodate}.
Closest to our paper is the literature on algorithmic collusion and AI-driven trading, which shows that adaptive agents can sustain supra-competitive outcomes and distort prices without explicit coordination \citep{calvano_artificial_2020,dou2025ai}.
Empirical results suggest that while RL is a standard framework for trading \citep{hamblyRecentAdvancesReinforcement2021}, agents may learn manipulative tactics to maximize rewards \citep{martinez-mirandaLearningUnfairTrading2015}. Real-world data shows this unguided optimization usually distorts markets, since algorithms exploit minor initial mispricing and amplify it into large asset bubbles \citep{zhangMispricingAlgorithmTrading2025}.
%
%
See \cref{app:related_works} for an extended discussion.

\section{Model}\label{sec:model}

We work in discrete time. At each time $t\ge1$, a trade of signed size $Q_t$ (positive for buys and negative for sells) is executed and produces a price impact $\delta_t$. Between $t$ and $t+1$, the market is hit by an exogenous shock $\noise_{t+1}$. The market price is the result of these two forces as defined below.
\begin{definition}[Market price]\label{def:priceformation}
Fix an initial price $P_1>0$. The market price $(P_t)_{t>1}$ is
\begin{equation}
\label{eq:price}
    P_{t+1} = (P_t + \delta_t) \noise_{t+1} \, ,
\end{equation}
where $\delta_t\in\mathbb{R}$ is the impact from trade and $\noise_{t+1}>0$ is the exogenous shock between time $t$ and $t+1$.
\end{definition}

We remark that the additivity of $\delta_t$ is standard in market microstructure~\citep{kyle_continuous_1985} while the multiplicativity of $\noise_{t+1}$ follows the standard in asset pricing~\citep{fama1970efficient}. We make the common assumption that the shocks are independent and identically distributed with finite variance, and normalize their mean to one.

\begin{assumption}
    \label{ass:noise}
    $(\noise_{t})_t$ is an i.i.d.\ stochastic process such that $\noise_t>0$ has finite variance and $\E{\noise_t} = 1$.
\end{assumption}



We next specify price impact. To obtain analytical results, we fix a functional form, although we do not expect the qualitative mechanism to depend on this choice as long as buy orders move prices up and sell orders move prices down. We adopt a square-root specification, following a large empirical literature on market impact that finds a square-root dependence on trade size~\citep{lillo2003master,toth2011anomalous,mastromatteo2014agent,toth2016square,bouchaud2018trades}.

\begin{assumption}
\label{ass:concaveprice}
Let $Q_t$ be the quantity traded at time $t$. Fix $\alpha_t \ge 0$ and $\beta_t \le 0$. The price impact is
\begin{equation}
\label{eq:delta}
    \delta_t = \begin{cases}
        \alpha_t \sqrt{Q_t} & Q_t \ge 0 \\
        \beta_t \sqrt{-Q_t} & Q_t < 0
    \end{cases}
\end{equation}
\end{assumption}
We note that $\alpha_t$ and $\beta_t$ represent the illiquidity of the market and summarize the maker’s liquidity provision that control the sensitivity of prices to the order flow. Illiquidity is defined in the sense of \citet{black1971toward}. When $\alpha_t = \beta_t = 0$, we have a perfectly liquid market where the price impact $\delta_t$ vanishes, implying that any trade size can be executed without affecting the price. In this limit, the price evolution is driven solely by the shocks $\noise_{t+1}$. When $\alpha_t$ or $\beta_t$ are large, we have an illiquid market where even moderate trade sizes can substantially move the price.


\subsection{Two-Player game}

We consider a repeated game between two representative agents: a market maker ($\maker$) and a market taker ($\taker$). At the beginning of the first round, the market price is $P_1$ and the maker (taker) is initialized with a non-negative amount of cash $\cash^\maker_1$ ($\cash^\taker_1$) and inventory $\inv^\maker_1$ ($\inv^\taker_1$). 
On every round $t\ge1$, the maker chooses illiquidity parameters $\alpha_t$ and $\beta_t$.
The taker subsequently decides the quantity $Q_t$ to trade.
The trade causes a price impact according to \cref{eq:delta} and the players exchange the quantity $Q_t$ for an amount of cash equal to $Q_t(P_t + \delta_t)$. This condition means that $P_t + \delta_t$ is the average trade price, which is similar to the fair pricing condition of~\citet{farmer2013efficiency}.
Finally, the price $P_{t+1}$ from \cref{eq:price} is revealed. 

Notice that, since $\noise_{t+1}$ has mean one, the average trade price equals the expected next price, so the maker's expected profit from trade execution is zero: $\mathbb{E}\!\left[Q_t(P_t+\delta_t)-Q_tP_{t+1}\mid P_t,\delta_t,Q_t\right]=0$. Thus, this fair-pricing condition represents the competitive limit in which many liquidity providers compete away expected profits from trade execution~\citep{biais2000competing}. 
Also notice that, by construction, the total amounts of inventory and cash are constant at every round. Thus, we define the constants $\inv=\inv^\maker_t + \inv^\taker_t$ and $\cash=\cash^\maker_t + \cash^\taker_t$. The game is summarized in \cref{tp:two-player-game}. 


Adaptive market agents are typically designed to maximize mark-to-market portfolio value: cash plus the market value of inventory at prevailing prices.
We model agents accordingly. Specifically, we define the wealth on round $t$ of any player $p \in \{\maker, \taker\}$ as the mark-to-market value of the player's portfolio
$
    \wealth^p_t = \cash^p_t + P_t \inv^p_t
$
and the objective of each player $p$ is to maximize the expected value of $\wealth^p_{t+1}-\wealth^p_{t}$ (see \cref{sec:learnbabylearn}) or the expected value of their long-run wealth as $T \to \infty$ (see \cref{sec:farsighted}).

\begin{algorithm}[t]
    \renewcommand*{\algorithmcfname}{Trading Protocol}
    \DontPrintSemicolon
    
    \KwData{
        Starting positions $(\inv^\maker_{1}, \cash^\maker_{1}) \ge 0$ and $(\inv^\taker_{1}, \cash^\taker_{1}) \ge 0$.
        Initial price $P_1 \ge 0$.
    }

    \For{round $t = 1, 2, \dots$} {
        Maker publishes $\alpha_t \ge 0$ and $\beta_t \le 0$\;
        Taker picks $Q_t \in \mathbb{R}$\;
        Price impact $\delta_t$ is computed using \cref{eq:delta}\;
        Taker updates inventory $\inv^\taker_{t+1} \gets \inv^\taker_t + Q_t$ and cash $\cash^\taker_{t+1} \gets \cash^\taker_t - Q_t (P_t+\delta_t)$\;
        Maker updates inventory $\inv^\maker_{t+1} \gets \inv^\maker_t - Q_t$ and cash $\cash^\maker_{t+1} \gets \cash^\maker_t + Q_t (P_t+\delta_t)$\;
        Price $P_{t+1}$ from \cref{eq:price} is revealed\;
    }

    \caption{Two-player game between \emph{maker} ($\maker$) and \emph{taker} ($\taker$).}
    \label[tp]{tp:two-player-game}
\end{algorithm}

\subsection{Strategy profiles}\label{sec:strategy}

The game defined in \cref{tp:two-player-game} is a general-sum Markov game~\citep{shapley1953,littman94} where the payoff is defined by the increase in wealth. We are interested in stationary \emph{Markov strategies} for the game, defined for any player $p \in \{\maker, \taker\}$ at any round $t$ as a map $\strat^p: \mathcal{S} \to \mathcal{P}(\mathcal{A}_p)$, where $\mathcal{S}$ is the state space, in our case consisting of the amounts of cash and inventory of both players and the price, and $\mathcal{P}(\mathcal{A}_p)$ is the set of all distributions over the action space of player $p$. A \emph{strategy profile} $\strat = (\strat^\maker, \strat^\taker)$ is defined as a pair of strategies, one for the taker and one for the maker.
A fundamental property of the strategy profiles we are interested in is \emph{price positivity}, which describes profiles that keep the market price strictly positive.
\begin{definition}[Price positivity]\label{def:pricepositivity}
    A strategy profile $\strat$ is price-positive if for all $t$, it holds that 
    $
        P_t > 0
    $
    almost surely with respect to the (possible) internal randomization of $\strat$ and the shocks $(\noise_t)_t$.
\end{definition}
We characterize price-positive strategy profiles within the model as follows.
\begin{restatable}{lemma}{pricepositivitycharacterization}[Price positivity characterization]\label{lemma:pricepositivity}
    A strategy profile $\strat$ is price-positive if and only if for all $t \ge 1$ such that $Q_t < 0$ it holds
    $
        \beta_t > \nicefrac{-P_t}{\sqrt{-Q_t}}
    $.
\end{restatable}

\subsubsection{Feasible profiles}

Here we are interested in \emph{feasible} strategy profiles that rule out short and leveraged positions.
\begin{definition}[Feasible strategy profile]\label{def:feasible-strategy}
    A strategy profile $\strat$ is feasible if it is price-positive and there exists a pair $(c, i) \in [0, \infty)^2$ such that, for all $t$ and for both players $p \in \{\maker, \taker\}$, it holds that 
    $
        \cash^p_t > c
    $ and $
        \inv^p_t > i
    $
    almost surely with respect to the (possible) internal randomization of $\strat$ and the shocks $(\noise_t)_t$.
\end{definition}
As for price positivity, we provide a characterization of feasible strategy profiles as a set of inequalities ensuring that the traded inventory and cash after each trade never exceed the players' reserves.
\begin{restatable}{lemma}{feasibilitycharacterization}[Feasibility characterization]\label{lemma:feasibility}
    A price-positive strategy profile is feasible if and only if there exists a pair $(c, i) \in [0, \infty)^2$ such that, for all $t$ the following set of inequalities holds:
    \begin{align}
        Q_t(P_t+\delta_t) < \cash^\taker_t - c
        \quad & \text{for} \quad
        Q_t \ge 0
        \label{eq:feasibility:1}
    \\
        Q_t < \inv^\maker_t - i
        \quad & \text{for} \quad
        Q_t \ge 0
        \label{eq:feasibility:2}
    \\
        -Q_t(P_t+\delta_t) < \cash^\maker_t - c
        \quad & \text{for} \quad
        Q_t < 0
        \label{eq:feasibility:3}
    \\
        -Q_t < \inv^\taker_t - i
        \quad & \text{for} \quad
        Q_t < 0
        \label{eq:feasibility:4}
    \end{align} 
\end{restatable}

\subsubsection{Overpricing profiles}

Here we are interested in strategy profiles that sustain prices above fundamentals. We define the fundamental price process $(F_t)_t$ as the counterfactual price path obtained by shutting down price impact ($\delta_t=0$) in \cref{def:priceformation}. In this counterfactual, prices respond only to the sequence of shocks $(\noise_t)_t$, so $F_t$ isolates the exogenous component of the price.

\begin{definition}[Fundamental price]\label{def:fundamental}
Let $P_1$ be the initial market price. Fix a sequence of realizations of the shocks $(\noise_t)_t$.
The fundamental price is $F_1 \coloneq P_1$ and $F_{t+1} \coloneq F_t\noise_{t+1}$ for $t>1$.
\end{definition}

Next, we call \emph{mispricing} the log difference between the market and fundamental prices, and we define a strong notion of persistent mispricing.

\begin{definition}[Mispricing]\label{def:mispricing}
    Let $P_t^\strat$ be the market price at time $t$ under  a price-positive strategy profile $\pi$. Let $F_t$ be the corresponding fundamental price. The mispricing is
    \begin{equation}
        M_t^\strat = \log P_t^\strat - \log F_t,
    \end{equation}
    and the mispricing drift is
    \begin{equation}
        m_t^\strat = \E{M_{t+1}^\strat - M_t^\strat}, 
    \end{equation}
    where the expectation is taken with respect to the internal randomness of $\strat$ and $(\noise_t)_t$.
\end{definition}

\begin{definition}[Persistent mispricing]
    \label{def:overpricing}
    A strategy profile $\strat$ generates persistent mispricing if
    \begin{equation}
        \Pb{\lim_{t\to\infty}  |M_t^\strat| = \infty} = 1,
    \end{equation}
    where the probability is taken with respect to the internal randomness of $\strat$ and $(\noise_t)_t$. In particular if $M_t^\strat \to \infty$ a.s., then we call this persistent \textit{overpricing}, while if $M_t^\strat \to -\infty$ a.s., then we call this persistent \textit{underpricing}.
\end{definition}


\subsubsection{Parameterization}



We parameterize the strategy profiles as follows.

\begin{strategy}\label{strat:phi}
Define a strategy profile $\strat$ with the hyper-parameters $\phi \in [0, 1]$ and $(c, i) \in [0, \infty)^2$, and the parameters $(\ka, \kb, \va, \vb) \ge 0$. 
At each time step $t$, the illiquidity parameters $(\alpha_t, \beta_t)$ and the traded quantity $Q_t$ are picked as
\[
	\alpha_t = \va \frac{P_t}{\sqrt{A_t}},
\quad
	\beta_t = - \vb \frac{P_t}{\sqrt{B_t}}
\qquad
\text{and}
\qquad
    Q_t = \begin{cases}
        + \ka^2 A_t & \text{w.p. } \ \phi \\
        - \kb^2 B_t & \text{w.p. } \ 1-\phi \\
    \end{cases},
\]
where
\[
    A_t = \min\lcb{\inv^\maker_t - i, \dfrac{\cash^\taker_t - c}{P_t} },
    \qquad
    B_t = \min\lcb{\dfrac{\cash^\maker_t - c}{P_t}, \inv^\taker_t - i}.
\]
\end{strategy}
The non-negativity condition on the parameters is motivated by \cref{ass:concaveprice} for the maker's parameters ($\va, \vb$) and w.l.o.g.\ for the taker's. Intuitively, the probability $\phi$ models the taker's propensity to buy or sell, while $A_t$ and $B_t$ are endogenous capacity constraints that represent the maximal tradable quantities without being forced into a short position and keeping the assets above the thresholds $i$ and $c$.
\Cref{strat:phi} provides a scale-free representation of feasible buy and sell actions: since $A_t$ and $B_t$ are the maximal quantities allowed by the cash and inventory constraints, any feasible buy or sell quantity can be written as a fraction of $A_t$ or $B_t$. State dependence enters through $A_t$, $B_t$, and $P_t$, while the parameters $(\ka, \kb, \va, \vb)$ capture stationary behavior after factoring out the endogenous scale induced by portfolio constraints.

Next, we characterize the region of the parameter space that identifies the set of feasible profiles.

\begin{restatable}{theorem}{feasible}\label{th:feasible}
    For any $\phi \in [0, 1]$ and any pair $(c, i) \in [0, \infty)^2$, a strategy profile $\strat$ is feasible if and only if
    $
        \ka < f_\alpha(\va)
    $ and $
        \kb < f_\beta(\vb)
    $,
    where $f_\alpha, f_\beta: \mathbb{R}^+ \to (0, 1]$ are well defined and decreasing.

\end{restatable}
The resulting feasibility boundaries $f_\alpha$ and $f_\beta$ highlight that, as illiquidity rises, traded quantities must shrink to ensure feasibility.


Finally, we provide a necessary and sufficient condition for a feasible strategy profile to generate persistent overpricing. We start by introducing an \emph{overpricing strength coefficient}.
\begin{restatable}{lemma}{mispricingcharacterization}
\label{lem:mispricingcharacterization}
    Under \cref{strat:phi}, the mispricing drift from \cref{def:mispricing} can be written solely as a function of the parameters
    \begin{equation}\label{eq:mu_overpricing}
        \mu^\strat \coloneq m^\strat_t
    =
        \phi \log(1 + \va \ka) + (1 - \phi) \log(1 - \vb \kb)
    \qquad \forall t
    \end{equation}
\end{restatable}
The next result shows how $\mu^\strat$ defines the region of the parameter space containing feasible strategy profiles that generate persistent overpricing.
\begin{restatable}{theorem}{overpricing}\label{th:overpricing}
    Any feasible stationary strategy profile \(\pi\) generates persistent overpricing if \(\mu^\pi>0\). It generates persistent underpricing if \(\mu^\pi<0\). If \(\mu^\pi=0\), then \(\pi\) generates neither persistent overpricing nor persistent underpricing.
\end{restatable}

In summary, \cref{strat:phi} defines the set of all strategy profiles $\Pi$. We characterized price-positive $\Pi_\mathrm{price-positive}$ and feasible $\Pi_\mathrm{feasible}$ strategy profiles, such that $\Pi_\mathrm{feasible} \subset \Pi_\mathrm{price-positive} \subset \Pi$, where all subsets are proper. We are interested in strategy profiles which are feasible and generate persistent overpricing $\Pi_\mathrm{overpricing} \cap \Pi_\mathrm{feasible}$. Note that this intersection is not empty by \cref{th:overpricing}.

\section{Strategic decomposition and learning dynamics in the myopic case}
\label{sec:learnbabylearn}

We are interested in a strategic taker and maker that, on each round, update the parameters of their respective strategies to maximize the immediate expected wealth increase $\Et{\wealth^p_{t+1} - \wealth^p_t}$, for any player $p \in \{\maker, \taker\}$ and any round $t$, where the expectation is taken with respect to the randomization of the taker's strategy, the shocks $\noise_{t+1}$ and conditioning on the history up to time $t$. Note that we can write the objective as
$
	\Et{\wealth^p_{t+1} - \wealth^p_t}
=
	\Et{P_{t+1} - P_t} \inv^p_t
$
using the update rules from \cref{tp:two-player-game,ass:noise}.
Now introduce $\kappa$ as the expected price impact normalized by the current price
\begin{equation}\label{eq:kappaexplicit}
	\kappa
\coloneq
    \frac{\Et{\delta_t}}{P_t}
=
    \phi \va \ka - (1 - \phi) \vb \kb \,,
\qquad \text{where} \quad
    \delta_t = \begin{cases}
        + \va \ka P_t & \text{w.p.} \quad \phi \\
        - \vb \kb P_t & \text{w.p.} \quad 1-\phi
    \end{cases}
\end{equation}
and the definition of $\delta_t$ comes from \cref{eq:stratdelta} and is a consequence of \cref{strat:phi}. Note that $\kappa$ depends only on the parameters of the players' strategy profile and is independent of the state. Using this definition, we can write the expected price difference as
\begin{align*}\label{eq:expecteddelta}
	\Et{P_{t+1} - P_t}
&= \tag{\cref{eq:price}}
	\Et{(P_t + \delta_t) \noise_{t+1} - P_t}
\\ &= \tag{Independence of $\noise_{t+1}$}
	\Et{\noise_{t+1}} P_t + \Et{\delta_t} \Et{\noise_{t+1}} - P_t
\\ &= \tag{\cref{eq:kappaexplicit}}
	\Et{\noise_{t+1}} P_t + \kappa \, \Et{\noise_{t+1}} P_t - P_t
\\ &= \tag{\cref{ass:noise}}
	\kappa P_t \,,
\end{align*}
for any round $t$. We can then write the immediate expected wealth increase solely as a function of the parameters
$
    R^p_t
\eqcolon
    \Et{P_{t+1} - P_t} \inv^p_t
=
	\kappa P_t \inv^p_t
$.
We are now able to formally define the \emph{one-shot} game on any round $t$ as follows.
\begin{game}[R]\label{game:R}
We define the sequential one-shot general-sum game played by taker and maker on every round $t$ on the stochastic game defined in \cref{tp:two-player-game}. The maker first picks parameters $(\va , \vb) \in [0, \infty)^2$ in the feasible region (\cref{th:feasible}), the taker responds by picking the parameters $(\ka, \kb) \in [0, f_\alpha(\va)) \times [0, f_\beta(\vb))$ in the feasible region.
The maker's utility is $R^\maker_t$, while the taker's utility is $R^\taker_t$.
\end{game}
Next, we decompose the reward of \gameref{game:R} into a competitive and collaborative component.

\subsection{Competitive game}
Consider the game where the players maximize the following utility
\begin{equation}\label{eq:utilitycompetitive}
	Z^\taker_t
\coloneq
	\Et{P_{t+1} - P_t} (\inv^\taker_t - \inv^\maker_t)
\quad \text{and} \quad
	Z^\maker_t
\coloneq
	\Et{P_{t+1} - P_t} (\inv^\maker_t - \inv^\taker_t)
\end{equation}
note that $Z^\taker_{t+1} = - Z^\maker_{t+1}$ and we can write it as
$
	Z^\taker_t
=
	-Z^\maker_t
=
	\kappa P_t (\inv^\taker_t - \inv^\maker_t)
$.
Next, define the \emph{competitive one-shot} game.
\begin{game}[Z]\label{game:Z}
We define the sequential one-shot zero-sum game played by taker and maker on every round $t$ on the stochastic game defined in \cref{tp:two-player-game}. The maker first picks parameters $(\va , \vb) \in [0, \infty)^2$ in the feasible region (\cref{th:feasible}), the taker responds by picking the parameters $(\ka, \kb) \in [0, f_\alpha(\va)) \times [0, f_\beta(\vb))$ in the feasible region.
The maker's utility is $Z^\maker_t$, while the taker's utility is $Z^\taker_t$ .
The game is zero-sum as $Z^\maker_t = - Z^\taker_t$.
\end{game}
Next, we show that the equilibrium point of this game defines a strategy profile $\strat_0$ such that $\delta_t = 0$ for all $t$, which implies that $P^{\strat_0}_t=F_t$ and there is no mispricing under $\strat_0$.
\begin{restatable}{theorem}{zoptimumisdeltazero}\label{th:zoptimumisdeltazero}
    Any strategy profile with no price impact ($\delta_t = 0$) is a Nash equilibrium for \gameref{game:Z}.
\end{restatable}

\subsection{Collaborative game}
Now introduce the game where the players optimize the utility
\begin{equation}\label{eq:utilitycollaborative}
	U^p_t \coloneq \Et{P_{t+1} - P_t} \inv \,.
\end{equation}
As both players have the same utility ($p$ does not appear on the right-hand side), we simply write $U_t$, which can be written as 
$
    U_t
=
	\kappa P_t \inv
$. Next, define the \emph{collaborative one-shot} game.
\begin{game}[U]\label{game:U}    
We define the sequential one-shot game played by taker and maker on every round $t$ on the stochastic game defined in \cref{tp:two-player-game}. The maker first picks parameters $(\va , \vb) \in [0, \infty)^2$ in the feasible region (\cref{th:feasible}), the taker responds by picking the parameters $(\ka, \kb) \in [0, f_\alpha(\va)) \times [0, f_\beta(\vb))$ in the feasible region.
The utility of both players is $U_t$, thus the game is purely potential~\citep{monderer1996potential}.
\end{game}
Call $W_t = W^\taker_t + W^\maker_t = C + P_t I$ the total wealth of the two players and note that it is proportional to the potential function of \cref{game:U}, as
\begin{align*}
    \E{W_t - W_1}
=
    \sum_{s=1}^{t-1} \E{\cash + P_{s+1} \inv} - \E{\cash + P_{s} \inv}
=
    \sum_{s=1}^{t-1} \E{P_{s+1} - P_s} \inv
=
    \sum_{s=1}^{t-1} U_s \,,
\end{align*}
where $W_1$ is a constant and the expectation is taken with respect to the (possible) internal randomization of $\strat$ and the shocks $(\noise_t)_t$.
The next result shows that the wealth of the two players under a strategy profile that generates persistent overpricing eventually dominates non-overpricing profiles.
\begin{restatable}{theorem}{uoptimumgeneratesoverpricing}\label{th:uoptimumgeneratesoverpricing}
    For total inventory $I>0$, let $\strat$ be any feasible stationary profile generating persistent overpricing, and let $\pi'$ be a feasible stationary profile that does not generate persistent overpricing. Then, along the same fundamental shock path,
    \[
        \Pb{
            \exists t_0<\infty \text{ such that }
            W_t^\pi > W_t^{\pi'}
            \text{ for all } t\ge t_0
        } = 1.
    \]
\end{restatable}


\subsection{Game decomposition and strategic equivalence}

We can decompose the utilities $R^\taker_t$ and $R^\maker_t$ of \gameref{game:R} into a fully competitive component (\gameref{game:Z}) and fully collaborative component (\gameref{game:U}) as
$
	R^\taker_t
=
	\frac{1}{2} Z^\taker_t
    + \frac{1}{2} U_t
$ and $
	R^\maker_t
=
	\frac{1}{2} Z^\maker_t
    + \frac{1}{2} U_t \,,
$
for any $\phi$ and any round $t$.

A pair of two-player games is strategically equivalent~\citep{maschler_game_2013, monderer1996potential} if the utilities of the two games coincide up to the scaling with a positive constant. \Gameref{game:R} is strategically equivalent to the \Gameref{game:U}:
for every fixed history up to round $t$, the induced one-shot \gameref{game:R} is strategically equivalent to \gameref{game:U}: for each player $p \in \{M,T\}$, $R_t^p = \frac{I_t^p}{I} U_t$, where $I_t^p / I > 0$ by feasibility when $I > 0$. Since the multiplicative factor is positive and independent of the player’s current action, the two one-shot games have the same best responses.
Therefore the best-response incentives of the original myopic game coincide with those of the collaborative component. By \cref{th:uoptimumgeneratesoverpricing}, profiles with positive mispricing drift $\mu^\pi>0$ eventually dominate profiles that do not generate persistent overpricing in total wealth. Hence, in positive-net-supply markets, myopic portfolio maximization induces incentives toward the overpricing region.

The stability of the competitive zero-impact solution ($\delta_t = 0$) in \gameref{game:Z} (\cref{th:zoptimumisdeltazero}) provides a counterweight to the result of \gameref{game:U} (\cref{th:uoptimumgeneratesoverpricing}): persistent overpricing is not an inevitable outcome of algorithmic trading, but rather contingent on the specific structure of the incentives, and this distinction becomes sharpest when we consider markets where the aggregate inventory is zero ($I=0$), which is characteristic of, for instance, prediction markets, where every \textit{long} position is matched by a \textit{short} position. In this regime, the collaborative component of the utility function vanishes ($U_t = 0$) and the interaction collapses into the purely competitive \gameref{game:Z}. Consequently, our model predicts that learning agents in zero-net-supply markets will remain trapped in competitive equilibria with no mispricing as per \cref{th:zoptimumisdeltazero}.

\subsection{Learnability of strategy profiles}
\label{sec:learnability-profiles}

We now study whether simple learning dynamics tend to produce overpricing over time.  Rather than tying the argument to a particular update rule, we first consider a general stochastic randomized block-coordinate scheme.  

\begin{definition}[Overpricing strength]
For $\gamma>0$, a feasible strategy profile $\strat$ is said to have
overpricing strength of at least $\gamma$ if $\mu^\strat\ge\gamma$.
\end{definition}

\begin{algorithmdef}[Stochastic randomized block-coordinate scheme]
\label{def:stoch_block_scheme}
Fix $\phi\in(0,1)$ and an initial feasible profile $\strat^0=(\va^0,\ka^0,\vb^0,\kb^0)$.  At each time $t$, the algorithm selects the $\alpha$-block with probability $\phi$ and the $\beta$-block with probability $1-\phi$.  If the $\alpha$-block is selected, only $(\va,\ka)$ is updated; if the $\beta$-block is selected, only $(\vb,\kb)$ is updated.  Thus, for some possibly stochastic block maps $\mathcal A_t,\mathcal B_t$,
\[
    \strat^{t+1}=
    \begin{cases}
        (\mathcal A_t(\strat^t),\vb^t,\kb^t), & \text{with probability }\phi,\\
        (\va^t,\ka^t,\mathcal B_t(\strat^t)), & \text{with probability }1-\phi,
    \end{cases}
\]
and the resulting profile is assumed feasible almost surely.
\end{algorithmdef}

We next define a drift-plus-noise condition ensuring that the learning dynamics makes positive expected progress.
Let $\mathcal F_t$ denote the history generated by the initial condition, past block choices, and past update noise.  For a target strength $\gamma>0$, define the overpricing margin $q_t^\gamma\coloneq\mu^\strat_t-\gamma$ and its increment $\Delta_{t+1} \coloneq q_{t+1}^\gamma-q_t^\gamma \equiv \mu^\strat_{t+1}-\mu^\strat_t$.

\begin{definition}[Stochastic progress condition]
\label{def:stoch_progress}
Fix $\gamma>0$ and a buffer $b\ge0$.  The scheme in \cref{def:stoch_block_scheme} is $(\nu_{\gamma,b},\sigma_\gamma)$-progressive up to buffer $b$ if $\nu_{\gamma,b}>0$, $\sigma_\gamma<\infty$, and, for all $t$, $\E{\Delta_{t+1}\mid\mathcal F_t}\ge\nu_{\gamma,b}\mathbf 1\{q_t^\gamma<b\}$, while $\Delta_{t+1}-\E{\Delta_{t+1}\mid\mathcal F_t}$ is conditionally $\sigma_\gamma$-sub-Gaussian.
\end{definition}

\begin{restatable}[Finite-time stochastic convergence to overpricing strength $\gamma$]{theorem}{ustochblockconv}
\label{th:stoch_block_overpricing}
Fix $\phi\in(0,1)$, $\gamma>0$, and $b\ge0$.  Let $\strat^t$ evolve according to \cref{def:stoch_block_scheme}, and define $\tau_{\gamma,b}\coloneq\inf\{t\ge0:q_t^\gamma\ge b\}$ and $G_{\gamma,b}\coloneq(b-q_0^\gamma)_+$.  If the scheme is $(\nu_{\gamma,b},\sigma_\gamma)$-progressive up to buffer $b$, then, for every $T\ge1$ such that $\nu_{\gamma,b}T>G_{\gamma,b}$,
\[
    \Pb{\tau_{\gamma,b}>T}
    \le
    \exp\!\left(-\frac{(\nu_{\gamma,b}T-G_{\gamma,b})^2}{2\sigma_\gamma^2T}\right),
\]
with the convention that the right-hand side is zero when $\sigma_\gamma=0$.  Consequently, $\tau_{\gamma,b}<\infty$ almost surely.  Moreover, for every $\delta\in(0,1)$, with probability at least $1-\delta$,
\[
    \tau_{\gamma,b}
    \le
    \left\lceil
        \frac{2G_{\gamma,b}}{\nu_{\gamma,b}}
        +
        \frac{8\sigma_\gamma^2}{\nu_{\gamma,b}^2}\log\frac1\delta
    \right\rceil .
\]
In particular, taking $b=0$ gives finite-time convergence to overpricing strength $\gamma$, with $G_{\gamma,0}=(\gamma-\mu^\strat_0)_+$.  If, in addition, the scheme satisfies $\E{\Delta_{t+1}\mid\mathcal F_t}\ge0$ whenever $q_t^\gamma\ge0$, then after hitting any buffer $b>0$, it remains in the overpricing region for any horizon $H\ge1$ with conditional probability at least $1-\exp(-b^2/(2\sigma_\gamma^2H))$.
\end{restatable}

The proof is in \cref{app:proof_stoch_block_overpricing}.  The theorem generalizes a deterministic pathwise monotonicity argument through a stochastic drift condition: noisy updates may temporarily decrease overpricing strength, but positive conditional progress yields high-probability finite-time entry and buffered finite-horizon persistence under the additional stability condition.

\begin{remark}[Projected stochastic gradient ascent]
\label{rem:psga_extension}
The abstract progress condition can be verified for concrete learning rules.  In
\cref{app:psga}, we do this for projected stochastic gradient ascent (PSGA) on the
reduced objective $\tilde\kappa=\va\ka-\vb\kb$.  PSGA update selects the
$\alpha$-block with probability $\phi$ and runs
\[
    \va^{t+1}=\va^t+\eta_{\va}(\ka^t+\xi_{\va,t}),\qquad
    \ka^{t+1}=\Pi_{\alpha,\epsilon_\pi}\!\left(\va^{t+1},
    \ka^t+\eta_{\ka}(\va^t+\xi_{\ka,t})\right),
\]
while $\beta$-block is selected with probability
$1-\phi$, where
\[
    \vb^{t+1}=\vb^t-\eta_{\vb}(\kb^t+\xi_{\vb,t}),\qquad
    \kb^{t+1}=\Pi_{\beta,\epsilon_\pi}\!\left(\vb^{t+1},
    \kb^t-\eta_{\kb}(\vb^t+\xi_{\kb,t})\right) ,
\]
with $\epsilon_\pi\ge0$, 
$
    \Pi_{\alpha,\epsilon_\pi}(\va,z)
    \coloneq
    \Pi_{[0,f_\alpha(\va)-\epsilon_\pi]}(z),
$  and
$
    \Pi_{\beta,\epsilon_\pi}(\vb,z)
    \coloneq
    \Pi_{[0,f_\beta(\vb)-\epsilon_\pi]}(z) \,.
$
The appendix works in product coordinates $x_t=\va^t\ka^t$ and
$y_t=\vb^t\kb^t$, using the equivalent criterion
$\mu^\strat_t\ge\gamma \iff x_t\ge r_\gamma(y_t)$.  Thus, defining
$s_t^\gamma=x_t-r_\gamma(y_t)$, PSGA inherits the same finite-time
convergence guarantee whenever the projected noiseless step gives positive
slack drift and the stochastic perturbation is sub-Gaussian and does not
dominate that drift.
\end{remark}

\subsubsection{Numerical experiments}
\label{sec:learningsimulation}

\begin{figure}
    \centering
    \begin{subfigure}[t]{.3\linewidth}
        \centering
        \includegraphics[width=\linewidth, page=1]{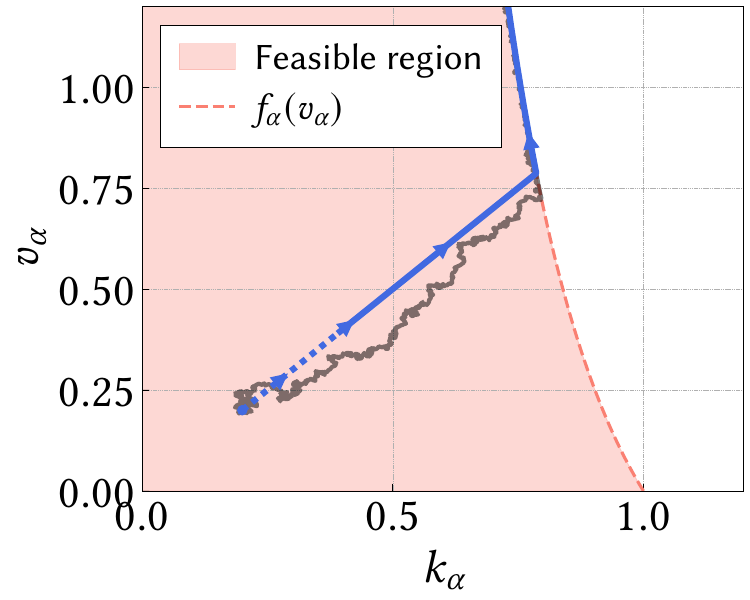}
        \caption{Ask parameters}
        \label{fig:learning-ask}
    \end{subfigure}\hfill
    \begin{subfigure}[t]{.3\linewidth}
        \centering
        \includegraphics[width=\linewidth, page=2]{learning-simulation.pdf}
        \caption{Bid parameters}
        \label{fig:learning-bid}
    \end{subfigure}\hfill
    \begin{subfigure}[t]{.3\linewidth}
        \centering
        \includegraphics[width=\linewidth, page=3]{learning-simulation.pdf}
        \caption{Joint parameter space}
        \label{fig:jointlearningagents}
    \end{subfigure}
    \caption{Plots \subref{fig:learning-ask} and \subref{fig:learning-bid} show a simulation of the learning trajectories of PSGA across the feasible space of parameters with $\phi = \nicefrac{1}{2}$.
    The dark line shows the learning trajectory obtained by noisy gradients, in cyan the noiseless trajectory. The dotted portion of the noiseless trajectory indicates profiles that generate no overpricing.
    Plot \subref{fig:jointlearningagents} shows a simulation of various learning trajectories of PSGA, the dark line shows the learning trajectory for $\phi = \nicefrac{1}{2}$, in cyan the average with respect to $\phi$.
    We highlight only one trajectory for clarity.
    }
    \label{fig:learningagents}
\end{figure}

Starting from a fixed point outside the overpricing region, we empirically simulate learning market agents that on every round update the parameters of the respective strategies using projected stochastic gradient ascent (see \cref{rem:psga_extension}).
That behavior on the ask side (\cref{fig:learning-ask}) is associated with high illiquidity and lowering trading volume, while the bid parameters (\cref{fig:learning-bid}) point to a region where $\vb \kb = 0$, associated with infinite liquidity and zero trading volume.
%
Next, for analogous learning agents, in \cref{fig:jointlearningagents} we simulated a grid of several starting configurations of the parameters within the feasible set and we show the actual iterates randomized by $\phi$
and the average trajectories with respect to $\phi$, both are clearly pointing inside the overpricing region regardless of the starting position.


\section{Farsighted objectives and connection to the myopic case}
\label{sec:farsighted}

In this section, we focus on agents that instead optimize a long-run criterion, namely the infinite-horizon expected wealth $\E{\wealth^p_T}$ as $T \to \infty$. Equivalently, for any player $p \in \{\maker,\taker\}$ and a profile $\strat$, define the farsighted objective as the infinite-horizon average log-wealth increase
\begin{equation}\label{eq:farsightedobjective}
	J^p_\strat \coloneq \lim_{T \to \infty} \frac{1}{T} \E{\log \wealth^p_T - \log \wealth^p_1},
\end{equation}
where the expectation is taken with respect to the internal randomization of profile $\strat$ and the shocks $(\noise_t)_t$.
\begin{assumption}\label{ass:boundednoise}
    Let $(\noise_t)_t$ be the i.i.d.\ shock process. There exists $d > 0$ s.t.\ $\E{ \noise_t^{d} } < \infty$ and $\E{ \noise_t^{-d} } < \infty$.
\end{assumption}
Under this assumption, the farsighted objective can be rewritten in a convenient way.
\begin{restatable}{lemma}{farsightedobjective}\label{lem:farsightedobjective}
    Under \cref{ass:boundednoise} and for any strategy profile $\strat$ such that $c, i > 0$, the farsighted objective can be written as $J^p_\strat = \max\{ \mu^\strat + \mu^\noise, 0 \}$ for all $p \in \{\taker, \maker\}$.
\end{restatable}


By the previous lemma, maximizing the farsighted objective over stationary
profiles is equivalent to maximizing \(\mu^\pi\). Thus the same parameter
regions favored by myopic incentives are also favored by the farsighted
objective. We now show that learning dynamics satisfying the progress condition
from \cref{th:stoch_block_overpricing} reach a farsighted near-optimal profile in finite time.
For any pair \((V_\alpha,V_\beta)>0\), define the constrained feasible region
\(X(V_\alpha,V_\beta)\) as
\[
    0 \le \va \le V_\alpha
\quad
    0 \le \vb \le V_\beta
\quad
    0 \le \ka < f_{\va}(\va)
\quad
    0 \le \kb < f_{\vb}(\vb)
\]
such that we can define
$
    \mu^\star
\coloneq
    \sup_{(\ka, \kb, \va, \vb) \in \mathcal X_{(V_\alpha, V_\beta)}} \mu^\strat(\ka, \kb, \va, \vb).
$

\begin{restatable}{theorem}{myopicoptimizesfarsighted}
\label{th:myopicoptimizesfarsighted}    
Under the assumptions of \cref{lem:farsightedobjective}, let \(J^\star:=\max\{\mu^\star+\mu^\varepsilon,0\}\). For any \(\epsilon>0\) with \(\mu^\star-\epsilon>0\), suppose \cref{th:stoch_block_overpricing} holds with target \(\gamma=\mu^\star-\epsilon\). Then the iterates of \cref{def:stoch_block_scheme} reach, in finite time almost surely, a profile \(\pi\) such that $J^p_\pi \ge J^\star-\epsilon$ for all $p\in\{M,T\}$.
\end{restatable}




\section{Conclusion}\label{sec:conclusion}

The purpose of the proposed model is not to predict that real-world prices will literally drift upward without bound, but to isolate an incentive alignment that can emerge when adaptive agents are evaluated on mark-to-market performance in markets with positive net supply. In this sense, the results should be read as a proof-of-concept instability: in the absence of countervailing forces that are deliberately abstracted away here, decentralized learning can converge to strategy regions that sustain persistent upward price pressure. Natural directions for future work include empirical tests of the distinction between positive-net-supply and zero-net-supply markets, as well as theoretical extensions with multiple makers and takers, heterogeneous learning rules, risk sensitivity, and alternative market designs or regulatory constraints.

\clearpage


\begin{ack}
EG acknowledges support by the Swiss National Science Foundation under Grant No. 233173.

AF is partially supported by project SERICS (PE00000014) under the NRRP MUR program funded by the EU - NGEU.

LF and NCB are partially supported by the MUR PRIN grant 2022EKNE5K (Learning in Markets and Society), by the FAIR (Future Artificial Intelligence Research) project, funded by the NextGenerationEU program within the PNRR-PE-AI scheme (M4C2, investment 1.3, line on Artificial Intelligence), and by the EU Horizon CL4-2022-HUMAN-02 research and innovation action under grant agreement 101120237, project ELIAS (European Lighthouse of AI for Sustainability).

TC acknowledges support by the FAIR (Future Artificial Intelligence Research) project, funded by the NextGenerationEU program within the PNRR-PE-AI scheme (M4C2, Investment 1.3, Line on Artificial Intelligence).

\end{ack}


\bibliographystyle{ACM-Reference-Format} 
\bibliography{references}

\clearpage


\appendix
\label{appendix}

\section{Related Work Discussion}\label{app:related_works}

This work contributes to several strands of literature. A first strand studies how market makers learn to optimize quoting and inventory-management rules. Early work connects market making to online convex optimization and develops no-regret guarantees under stylized feedback and execution models~\citep{10.1145/2465769.2465777,abernethy2013adaptive}, with more recent contributions emphasizing partial-information settings and richer trading frictions~\citep{cesa-bianchi_market_2024}. Related work examines robustness to shocks~\citep{10.5555/2981780.2981826} and the implications of algorithmic pricing for liquidity and market quality~\citep{colliard_algorithmic_2022}. Recently, the problem has been extended to the realm of decentralized finance, with studies focusing on the optimal design of constant function market makers and strategies for liquidity provision in platforms like Uniswap~\citep{bar-on_uniswap_2023}. An extensive body of experimental work has successfully applied reinforcement learning techniques to the market-making problem~\citep{10.5555/3237383.3237450, 10.5555/3398761.3399059,ganesh_reinforcement_2019}, using techniques from deep learning in high-frequency trading~\citep{pmlr-v189-kumar23a} and in the context of limit order books~\citep{wei_model-based_2019, coletta_learning_2022}. Unlike this literature, which focuses on designing a single agent for the market-making problem, we study the strategic interaction between a market maker and a market taker and ask whether it can endogenously generate price distortions.

A second strand studies online learning for trading and market taking, often formalized as online portfolio selection. Starting from the universal portfolio framework of~\citet{cover1,cover2}, this literature develops algorithms that compete with the best constant-rebalanced portfolio in hindsight and refines the computational and statistical foundations of such guarantees~\citep{892136,hazan_online_2015,zimmert_pushing_2022,jezequel_efficient_2025}. Further theoretical work has explored the connections between stochastic and worst-case models for investing~\citep{hazan_stochastic_2009,putta_data_2025}, providing a more comprehensive understanding of performance guarantees in different market settings. While these contributions deliver powerful performance guarantees for a single trader facing an exogenous price process, our setting is intrinsically strategic: the taker's trades affect prices through impact, and the resulting price dynamics feed back into both agents' wealth and future incentives.

A third strand concerns learning in strategic environments. Multi-agent reinforcement learning~\citep{albrecht_multi-agent_nodate} provides a natural toolkit for agent economies, but general convergence guarantees are scarce because each agent faces a non-stationary environment induced by others~\citep{daskalakis2009complexity}. Convergence has been shown only in special cases such as Q-learning~\citep{watkins1992} in two-player zero-sum games~\citep{littman94}, the iterated Prisoner's Dilemma~\citep{sandholm_crites_1995}, and more general arbitrary-sum two-player games assuming Nash equilibrium play~\citep{hu_wellman_1998}. Our contribution is to exploit the game's economic structure to recover tractability. We map the learning problem into a low-dimensional parameter space in which the mispricing regimes can be characterized explicitly, and we then study adaptive dynamics within this space.

Finally, our paper is tied to the emerging literature on algorithmic collusion among learning agents. While algorithmic collusion describes systems in which the repeated game effectively exhibits different equilibrium properties compared to the stage game, the connection between algorithmic collusion and overpricing is common in the literature \citep{calvano_artificial_2020,dou2025ai} since overpricing is an outcome of collusion. Early work shows that adaptive agents can learn to soften competition in repeated Cournot settings~\citep{waltman2008}, and subsequent work demonstrates that standard Q-learning and related methods can sustain collusive pricing in repeated Bertrand environments via implicit reward–punishment schemes~\citep{calvano_artificial_2020,harrington2018collusion}. Recent simulation evidence suggests analogous concerns can arise in financial settings, with AI-driven speculators exhibiting supra-competitive outcomes even absent explicit agreement~\citep{dou2025ai}. Related results also highlight the role of information and foresight in shaping the extent of collusion in dynamic auction environments~\citep{banchio_artificial_2022,banchio_adaptive_2023}, and a recent result by \citet{cartea_algorithmic_2026} proves a version of the Folk theorem for learning agents in a repeated potential game, showing that, under the right conditions, there is a positive probability of learning a collusive strategy profile. In contrast, the overpricing we study does not arise from a mismatch between the stage and repeated games, but from a structural incentive alignment. When aggregate inventory is positive, higher prices increase aggregate portfolio values, creating a collaborative component in otherwise non-collaborative objectives.

\section{Simulations}
\label{sec:marketsimulation}

\begin{figure}
\begin{subfigure}{\linewidth}
    \centering
    \includegraphics[width=\linewidth]{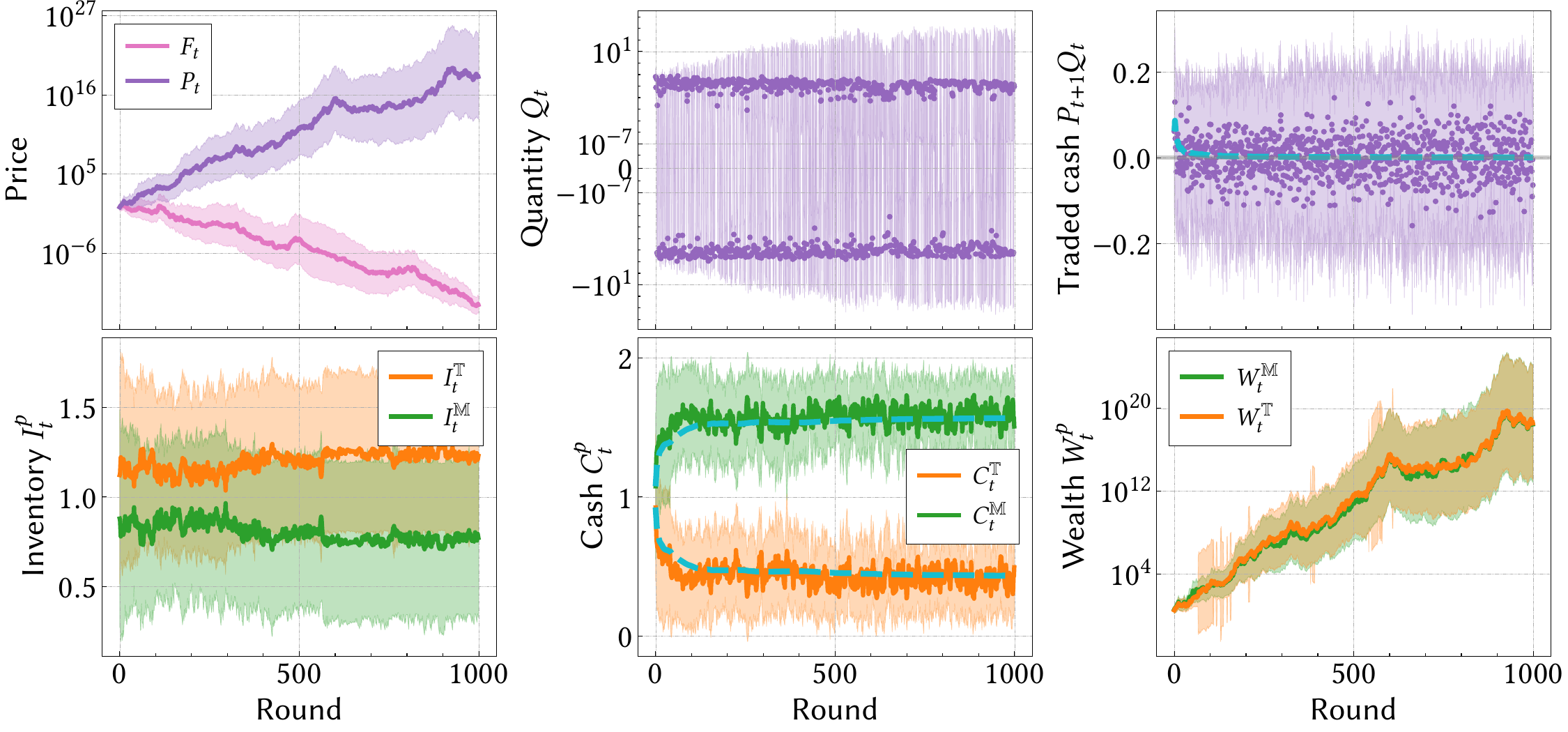}
    \caption{Market impact of a stationary overpricing profile $\strat^+$ with $\ka = \kb = \va = \vb = \nicefrac{1}{2}$ and $\phi = 0.7$.}
    \label{fig:sim-coll-strat}
\end{subfigure}\vspace{.3cm}
\begin{subfigure}{\linewidth}
    \centering
    \includegraphics[width=\linewidth]{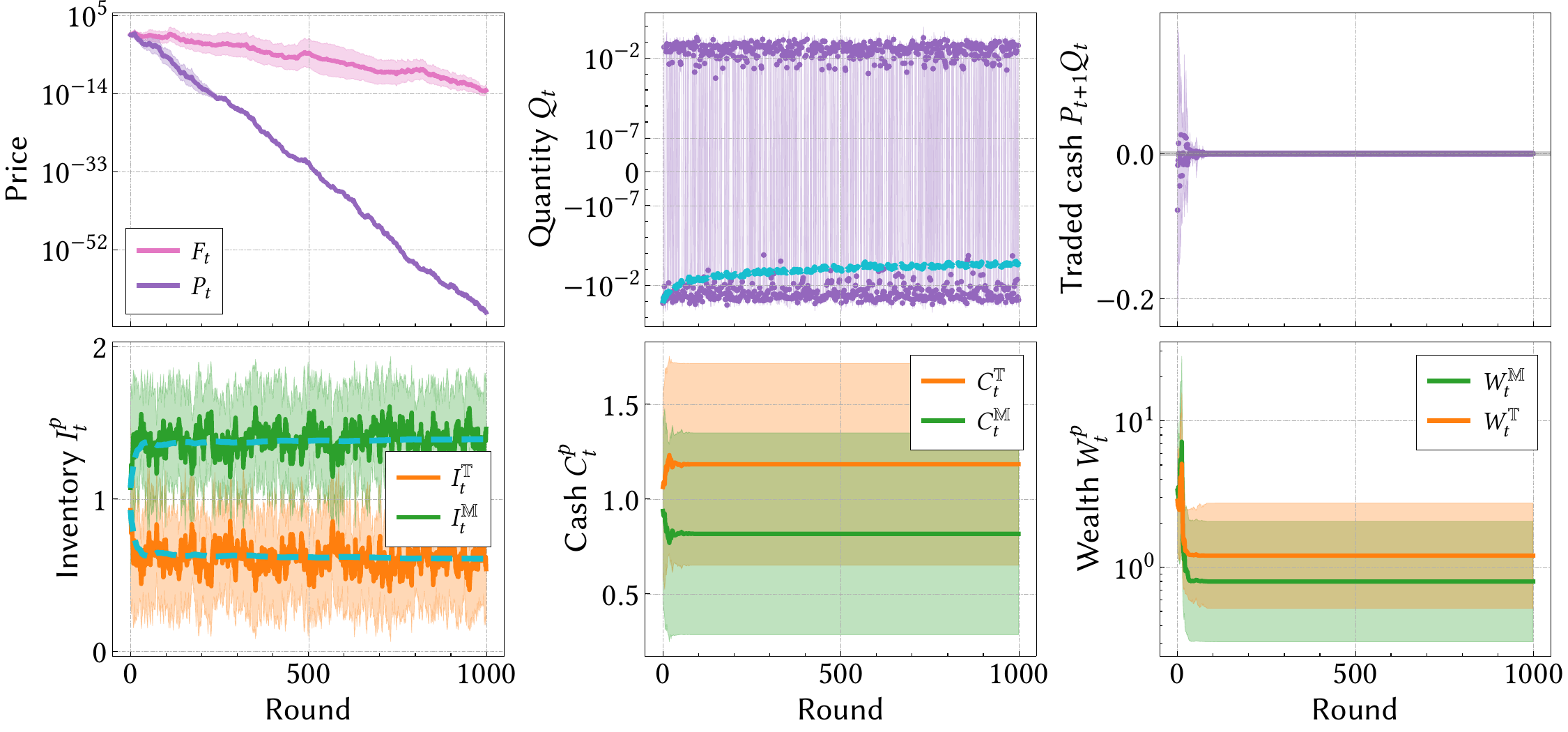}
    \caption{Market impact of a stationary underpricing profile $\strat^-$ with $\ka = \kb = \va = \vb = \nicefrac{1}{2}$ and $\phi = 0.3$.}
    \label{fig:sim-non-coll-strat}
\end{subfigure}
\caption{Simulations comparing side-by-side the impact on the market of an overpricing strategy profile $\strat^+$ (\cref{fig:sim-coll-strat}) and an underpricing one $\strat^-$ (\cref{fig:sim-non-coll-strat}) over 1000 trading days across 50 independent runs.
Each group of plots is split into a top row for public market signals (in purple): price, quantity, and traded cash, while the bottom row contains player-specific signals (orange for the taker and green for the maker): inventory, cash and wealth.
For clarity, the plots of the price, quantity and wealth are in (sym)log scale.
The faded regions represent the standard deviation across the runs around the average, which is plotted in a darker shade. The dashed cyan line indicates the running average.
See \cref{sec:marketsimulation} for an in-depth discussion.
}
\label{fig:simulation}
\end{figure}

To better understand the dynamics of the model from \cref{tp:two-player-game} and the strategy profiles in \cref{strat:phi}, we simulate the long-term effects of two stationary feasible strategy profiles on the price and the inventories and cash reserves of the players. We consider a profile generating persistent overpricing and one generating persistent underpricing, we ignore the case where the profile does not generate mispricing as the price $P_t$ would track the fundamental $F_t$ in probability. The players trade for $1000$ days across $50$ independent runs, averaging the results and showing standard deviation. The feasibility constraints are set to $c = i = 0$ and $C = I = 1$. At the beginning of each run, a random initial price $P_1$ is drawn from a log-normal distribution with mean zero and unit standard deviation, the players are endowed with a randomized initial amount of inventory and cash, both picked uniformly at random within the feasible sets:
\[
    P_1 \sim \mathrm{LogNormal} (0, 1)
\qquad
    I^\taker_1 \sim U([i, I]) \quad I^\maker_1 = I - I^\taker_1
\qquad
    C^\taker_1 \sim U([c, C]) \quad C^\maker_1 = I - C^\taker_1
\]
We draw shocks as
\[
    \noise_t \sim \mathrm{LogNormal} \! \lrb{ -\frac{\sigma^2}{2}, \sigma^2 },
\]
so that $\E{\varepsilon_t}=1$ and we pick $\sigma = 0.3$. In this case $\mu^\varepsilon = \E{\log\varepsilon_t} = -\sigma^2/2$, and overpricing requires $\mu^\pi+\mu^\varepsilon>0$.
For each round $t$, we plot the average value and standard deviation of several market features across the repeated experiments. When appropriate, we also include a running average as a dotted cyan line.

Both strategies parameterized by $\ka = \kb = \va = \vb = \nicefrac{1}{2}$, setting $\phi = 0.7$ yields the overpricing strategy profile $\strat^+$ for which $\mu^\strat > 0$ and setting $\phi = 0.3$ yields the underpricing strategy profile $\strat^-$ for which $\mu^\strat < 0$. Note that the discussed long-term dynamics are robust to any choice of strategy profiles $(\strat^+, \strat^-)$ such that $\mu^\strat + \mu^\noise > 0$ and $\mu^\strat + \mu^\noise < 0$ respectively.
In \cref{fig:simulation} we show the results.

As expected, looking at the difference between the actual price $P_t$ and the fundamental price $F_t$ in the top left plots of \cref{fig:sim-coll-strat,fig:sim-non-coll-strat}, we can see strong overpricing under $\strat^+$ and strong underpricing under $\strat^-$. 

As defined in \cref{tp:two-player-game}, on every round the players exchange a quantity $Q_t$ and cash $(P_t + \delta_t) Q_t$. The simulations show that the traded quantity $Q_t$ converges to zero under $\strat^+$ (\cref{fig:sim-coll-strat}, top center), which happens because $Q_t \in (-B_t, A_t)$ and both $A_t$ and $B_t$ tend to zero a.s.\ at rate $1 / P_t$ (see their definition in \cref{strat:phi}). The exchanged cash on the other hand does not converge to any value (\cref{fig:sim-coll-strat}, top right). This difference is crucial because, even if the traded quantity vanishes, the cash exchanged during the trade remains stable, ensuring price growth. The opposite happens under $\strat^-$ (\cref{fig:sim-non-coll-strat}, top center and right): the quantity does not converge, while the traded cash converges to zero at rate $P_t$ because $(P_t + \delta_t) Q_t \in (-B_t P_t, A_t P_t)$. 

The behavior of the traded quantity $Q_t$ and traded cash $P_t + \delta_t$ reflects on the inventories $(\inv^\maker_t, \inv^\taker_t)$ and cash amounts $(\cash^\maker_t, \cash^\taker_t)$ of the players. The inventory is the sum of the traded quantities, therefore if it has a limit, it lies between $0$ and $I$ by feasibility. Under $\strat^+$, we already established that $Q_t$ converges to zero at a rate $1 / P_t$, where $P_t$ grows exponentially in time, therefore the inventory converges (\cref{fig:sim-coll-strat}, bottom left); the cash on the other hand does not converge (\cref{fig:sim-coll-strat}, bottom center).

In conclusion, the dynamics that result from the overpricing and underpricing strategy profiles are symmetric and show that the former leads the traded quantity to zero and together with the non-zero traded cash causes the price to diverge, while the latter leads the traded cash to zero and the quantity remains non-zero, causing the price to converge to zero.
Interestingly, these differences seem to disappear in the average: as shown by the cyan dotted lines, the running average of the traded cash under $\strat^+$, as well as the running average of the traded quantity under $\strat^-$, seem to converge to zero. A more in-depth theoretical analysis of these quantities is left for future developments.



\section{Technical appendix}

In this section, we present the remaining proofs of the results presented in the paper.

\subsection{Proof of \Cref{lemma:pricepositivity}}
\label{app:proofpricepositivity}

\pricepositivitycharacterization*

\begin{proof}
    The proof proceeds by induction, we have $P_1 > 0$ by construction. Then, for every $t \ge 1$ we assume that $P_t > 0$ and
    \begin{equation}\label{eq:priceposassumption}
        \beta_t > -\frac{P_t}{\sqrt{-Q_t}}.
    \end{equation}
    We want to show that $P_{t+1} > 0$, where $P_{t+1} = (P_t + \delta_t) \noise_{t+1}$.
    By \cref{def:priceformation}, $\noise_t > 0$ for all $t$, therefore $(P_t + \delta_t)\e_{t+1} > 0$ holds as long as $P_t + \delta_t > 0$.
    If $Q_t > 0$, then $\delta_t > 0$ by \cref{eq:delta}. Otherwise, we have that $P_t + \beta_t \sqrt{-Q_t} > 0$ by \cref{eq:priceposassumption}. Conversely, if \cref{eq:priceposassumption} is violated on any round $t^\star$, then $P_{t^\star} < 0$ on such round if $Q_{t^\star} < 0$.
\end{proof}

\subsection{Proof of \Cref{lemma:feasibility}}
\label{app:prooffeasibility}

\feasibilitycharacterization*
\begin{proof}
We show that if all the inequalities hold, then any price-positive strategy profile is feasible. First, by the assumption on price positivity (\cref{def:pricepositivity}), it holds that $P_t > 0$ for any $t$ almost surely.

The proof is by induction. At time $t = 1$ we have $\cash^p_1 > c$, $\inv^p_1 > i$ by construction (see \cref{tp:two-player-game}). Regarding the induction step, for every $t \ge 1$ such that $\cash^p_t > c$, $\inv^p_t > i$, we have $\cash^p_{t+1} > c$, $\inv^p_{t+1} > i$ as shown below:
\begin{itemize}
    \item For $Q_t \ge 0$:
    \begin{itemize}
        \item $\cash^\maker_{t+1} = \cash^\maker_t + Q_t (P_t + \delta_t) > c$ as $\cash^\maker_t > c$ and $P_t + \delta_t > 0$. 
        \item $\cash^\taker_{t+1} = \cash^\taker_t - Q_t (P_t + \delta_t) > c$ by \cref{eq:feasibility:1}.
        \item $\inv^\maker_{t+1} = \inv^\maker_t - Q_t > i$ by \cref{eq:feasibility:2}.      
        \item $\inv^\taker_{t+1} = \inv^\taker_t + Q_t > i$ as $\inv^\taker_t > i$. 
    \end{itemize}
    \item For $Q_t < 0$:
    \begin{itemize}
        \item $\cash^\maker_{t+1} = \cash^\maker_t + Q_t (P_t + \delta_t) > c$ by \cref{eq:feasibility:3}.
        \item $\cash^\taker_{t+1} = \cash^\taker_t - Q_t (P_t + \delta_t) > c$ as $\cash^\taker_t > c$ and $P_t + \delta_t > 0$.
        \item $\inv^\maker_{t+1} = \inv^\maker_t - Q_t > i$ as $\inv^\maker_t > i$.
        \item $\inv^\taker_{t+1} = \inv^\taker_t + Q_t > i$ by \cref{eq:feasibility:4}.
    \end{itemize}
\end{itemize}
    
Conversely, we show that if any of the inequalities is violated, then the strategy profile is not feasible. 
If \cref{eq:feasibility:1} is violated for $Q_{t^*} \ge 0$ then $\cash^\taker_{t^*+1} < c$. 
If \cref{eq:feasibility:2} is violated for $Q_{t^*} \ge 0$ then $\inv^\maker_{t^*+1} < i$. 
If \cref{eq:feasibility:3} is violated for $Q_{t^*} < 0$ then $\cash^\maker_{t^*+1} < c$. 
If \cref{eq:feasibility:4} is violated for $Q_{t^*} < 0$ then $\inv^\taker_{t^*+1} < i$. 
If Price positivity is violated for $Q_{t^*} < 0$ then $\cash^\taker_{t^*+1} < c$. 
\end{proof}

\subsection{Proof of \Cref{th:feasible}}
\label{app:prooffeasible}

\feasible*
\begin{proof}
    Pick any $\phi \in [0, 1]$ and pair $(c, i) \in [0, \infty)^2$, let $\strat$ be a strategy profile parameterized by $(\ka, \kb, \va, \vb) > 0$. By \cref{eq:delta} we have:
    \begin{equation}\label{eq:stratdelta}
        \delta_t = \begin{cases}
            + \va \ka P_t & \text{w.p.} \quad \phi \\
            - \vb \kb P_t & \text{w.p.} \quad 1-\phi \,.
        \end{cases}
    \end{equation}
    We require $\strat$ to be feasible. Assume that $P_1 > 0$, $\inv^\maker_1 > i$, $\inv^\taker_1 > i$, $\cash^\maker_1 > c$ and $\cash^\taker_1 > c$. Consider any round $t$.
    Price positivity is achieved as per \cref{lemma:pricepositivity} when $Q_t < 0$ if
    \begin{equation}\label{eq:nonegativeprice}
        \beta_t > -\frac{P_t}{\sqrt{-Q_t}}
    \iff
        - \vb \cdot \frac{P_t}{\sqrt{B_t}} > -\frac{P_t}{\kb^2 \sqrt{B_t}}
    \iff
        \vb \kb < 1
    \end{equation}
    
    The feasibility characterization from \cref{lemma:feasibility} implies that a strategy profile is feasible if and only if the following set of inequalities is satisfied:
    \begin{align}
        \ka^2 A_t(P_t+\va \ka P_t) < \cash^\taker_t - c
        & \quad \text{for} \quad
        Q_t \ge 0
        \label{eq:th1:1}
    \\
        \ka^2 A_t < \inv^\maker_t - i &
        \quad \text{for} \quad
        Q_t \ge 0
        \label{eq:th1:2}
    \\
        \kb^2 B_t(P_t- \vb \kb P_t) < \cash^\maker_t - c
        & \quad \text{for} \quad
        Q_t < 0
        \label{eq:th1:3}
    \\
        \kb^2 B_t < \inv^\taker_t - i
        & \quad \text{for} \quad
        Q_t < 0
        \label{eq:th1:4}
    \end{align}
    From \cref{eq:th1:1} we have:
    \begin{equation}\label{eq:th1:6}
        P_t \ka^2 A_t (1+\va \ka)
    \le
        \ka^2 (\cash^\taker_t - c) (1+\va \ka)
    <
        \cash^\taker_t - c
    \iff
        \ka^2 + \va\ka^3 < 1\,.
    \end{equation}
    From \cref{eq:th1:2} we have:
    \begin{equation}\label{eq:th1:7}
        \ka^2 A_t
    \le
        \ka^2 (\inv^\maker_t - i)
    <
        \inv^\maker_t - i
    \iff
        \ka < 1\,.
    \end{equation}
    From \cref{eq:th1:3} we have:
    \begin{equation}\label{eq:th1:8}
        P_t \kb^2 B_t (1- \vb \kb)
    \le
        \kb^2 (\cash^\maker_t - c) (1- \vb \kb)
    <
        \cash^\maker_t - c
    \iff
        \kb^2 - \vb\kb^3 < 1\,.
    \end{equation}
    From \cref{eq:th1:4} we have:
    \begin{equation}\label{eq:th1:9}
        \kb^2 B_t
    \le
        \kb^2 (\inv^\taker_t - i)
    <
        \inv^\taker_t - i
    \iff
        \kb < 1\,.
    \end{equation}
    Notice that \cref{eq:th1:8} is redundant because it is satisfied by any $\vb\ge0$ when $\kb\le 1$. Note that the constraints are oblivious to the feasibility parameters $(c, i)$, which allows us to consider profiles where $c = i = 0$ without loss of generality.
    The constraints obtained are
    \begin{equation}\label{eq:wrongorderconstraints}
        0 \le \va < \frac{1 - \ka^2}{\ka^3}
    \quad
        0 \le \vb < \frac{1}{\kb}
    \quad
        0 \le \ka < 1
    \quad
        0 \le \kb < 1
    \end{equation}
	By the definition of \cref{tp:two-player-game}, the maker picks their parameters before the taker, to reflect that we require a formulation equivalent to \cref{eq:wrongorderconstraints}, but of the form
	\[
		\va \ge 0
    \quad
		\vb \ge 0
    \quad
		0 \le \ka < f_\alpha(\va)
    \quad
		0 \le \kb < f_\beta(\vb)
	\]
	To find $f_\alpha$, recall \cref{eq:th1:6} and write the constraint as $\ka^3 \va + \ka^2 - 1 < 0$.
	For a fixed $\va \ge 0$, the function $g(\ka) = \ka^3 \va + \ka^2 - 1$ is strictly increasing in $\ka \ge 0$ and therefore there is only one value $f_\alpha(\va) \in [0, 1]$ such that $g(f_\alpha(\va)) = 0$. To find $f_\alpha(\va)$ we need to solve the equation $g(x) = x^3 \va + x^2 - 1 = 0$. Consider the variable swap $t = \nicefrac{1}{x}$, thus $g(\nicefrac{1}{t}) = t^3 - t - \va = 0$.
	We are interested in the root $t(\va) \ge 1$:
	\[
		t(\va)
	\coloneq
		\sqrt[3]{\frac{\va}{2} + \sqrt{\frac{\va^{2}}{4} - \frac{1}{27}}}
	+
		\sqrt[3]{\frac{\va}{2} - \sqrt{\frac{\va^{2}}{4} - \frac{1}{27}}}
	\]
	Finally, applying the variable swap again we get $f_\alpha(\va) \coloneq \nicefrac{1}{t(\va)}$. To find $f_\beta$, consider the region $\vb \ge 0$ from price positivity in \cref{eq:nonegativeprice} and the region $\kb < 1$ from the inventory constraint in \cref{eq:th1:9}, to we get $f_\beta(\vb) \coloneq \min\{1, \nicefrac{1}{\vb}\}$.
\end{proof}

The difference between $f_\alpha$ and $f_\beta$ is tied to price-positivity, which is implied by feasibility and is a constraint on the bid side forcing the traded quantity to be finite even in markets with infinite liquidity.
Note that, while $f_\alpha$ is decreasing in its argument, the function $\va \mapsto \va \cdot f_\alpha(\va)$ is increasing, this is clear from the lower bound $\va \cdot f_\alpha(\va) \ge \nicefrac{\va}{1 + \sqrt{\va}}$.

\subsection{Proof of \cref{lem:mispricingcharacterization}}
\label{app:proofofmispricingcharacterization}
\mispricingcharacterization*
\begin{proof}
    Fix any strategy profile $\strat$. By \cref{def:fundamental} and \cref{def:priceformation}, we can write
    \[
        \E{M^\strat_{t+1} - M^\strat_t}
        = \E{\log \lrb{\frac{P^\strat_{t+1}}{F_{t+1}} \cdot \frac{F_t}{P^\strat_t}}}
        = \E{\log \lrb{1 + \frac{\delta_t}{P^\strat_t}}}
    \]
    If $\strat$ is parametrized according to \cref{strat:phi}, then we can write the price impact from \cref{eq:delta} as
    \[
        \delta_t = \begin{cases}
            + \va \ka P_t & \text{w.p.} \quad \phi \\
            - \vb \kb P_t & \text{w.p.} \quad 1 - \phi \\
        \end{cases}
    \]
    and plugging this back into the previous display we get
    \[
        \E{\log \lrb{1 + \frac{\delta_t}{P^\strat_t}}}
        = \phi \log(1 + \va \ka) + (1 - \phi) \log(1 - \kb \vb).
        \qedhere
    \]
\end{proof}

\subsection{Proof of \Cref{th:overpricing}}
\label{app:proofoverpricing}

We first prove an auxiliary result for the case for $\mu^\strat = 0$.

\begin{lemma}
\label{lem:zerodriftoscillation}
Fix a stationary feasible strategy profile such that \(\mu^\pi=0\), with \(k_\alpha v_\alpha\neq0\) and \(k_\beta v_\beta\neq0\). Then, almost surely,
\[
    \limsup_{t\to\infty} M_t^\pi=+\infty,
    \qquad
    \liminf_{t\to\infty} M_t^\pi=-\infty .
\]
In particular, \(\pi\) does not generate persistent mispricing.
\end{lemma}

\begin{proof}
Since the profile is stationary, the multipliers
\[
    \eta_t:=1+\frac{\delta_t}{P_t}
\]
are i.i.d.\ and take two values:
\[
\eta_t=
\begin{cases}
1+\ka\va & \text{with prob. }\phi,\\
1-\kb\vb & \text{with prob. }1-\phi,
\end{cases}
\qquad
\phi\in(0,1),
\qquad
1-\kb\vb>0.
\]
Define
\[
    a:=\log(1+\ka\va)>0,
    \qquad
    b:=-\log(1-\kb\vb)>0.
\]
Let \(X_t:=\log\eta_t\in\{a,-b\}\) and
\(S_n:=\sum_{t=1}^n X_t\). Since \(\mu^\strat=0\),
\[
    \E{X_1}=\phi a-(1-\phi)b=0.
\]
Moreover, by definition of mispricing,
\[
    M_{n+1}^\strat
    =
    M_1^\strat+\sum_{t=1}^n\log\eta_t
    =
    S_n,
\]
as \(M_1^\strat=0\). Thus \((S_n)_n\) is a martingale with bounded increments.

Fix \(m,n\in\mathbb N\) and define
\[
    \tau:=\inf\{t\ge 1:\ S_t\ge m\ \text{or}\ S_t\le -n\}.
\]
For each \(N\), optional stopping gives
\[
    \E{S_{\tau\wedge N}}=0.
\]
Since the overshoots are bounded,
\[
    S_\tau\in[m,m+a]\quad\text{on }\{S_\tau\ge m\},
    \qquad
    S_\tau\in[-n-b,-n]\quad\text{on }\{S_\tau\le -n\}.
\]
Hence \(|S_{\tau\wedge N}|\le m+a+n+b\), so dominated convergence yields
\[
    \E{S_\tau}=0.
\]
Let \(p:=\Pb{S_\tau\ge m}\). Then
\[
    0=\E{S_\tau}
    \le
    p(m+a)+(1-p)(-n),
\]
so
\[
    p\ge \frac{n}{m+a+n}.
\]
Letting \(n\to\infty\) gives
\[
    \Pb{\sup_{t\ge1}S_t\ge m}=1
\]
for every \(m\), hence \(\limsup_{t\to\infty}S_t=+\infty\) a.s.

The lower tail is analogous. Define
\[
    \tau':=\inf\{t\ge1:\ S_t\le -m\ \text{or}\ S_t\ge n\}
\]
and let \(q:=\Pb{S_{\tau'}\le -m}\). Since
\[
    S_{\tau'}\ge -m-b
    \quad\text{on }\{S_{\tau'}\le -m\},
    \qquad
    S_{\tau'}\ge n
    \quad\text{on }\{S_{\tau'}\ge n\},
\]
we get
\[
    0=\E{S_{\tau'}}
    \ge
    q(-m-b)+(1-q)n,
\]
and therefore
\[
    q\ge \frac{n}{m+b+n}.
\]
Letting \(n\to\infty\) gives
\[
    \Pb{\inf_{t\ge1}S_t\le -m}=1
\]
for every \(m\), hence \(\liminf_{t\to\infty}S_t=-\infty\) a.s.

Since \(M_{t+1}^\strat=S_t\), we conclude
\[
    \limsup_{t\to\infty}M_t^\strat=+\infty,
    \qquad
    \liminf_{t\to\infty}M_t^\strat=-\infty
    \qquad\text{a.s.}
\]
Thus the profile generates no persistent directional mispricing.
\end{proof}

We are now ready to prove the main result.

\overpricing*

\begin{proof}
Fix a feasible stationary profile \(\pi\) under \cref{strat:phi}. Recall that
\[
    M_t^\pi=\log P_t^\pi-\log F_t .
\]
Since \(F_{t+1}=F_t\varepsilon_{t+1}\) and
\(P_{t+1}^\pi=(P_t^\pi+\delta_t)\varepsilon_{t+1}\), we have
\[
    M_{t+1}^\pi-M_t^\pi
    =
    \log\frac{P_{t+1}^\pi/F_{t+1}}{P_t^\pi/F_t}
    =
    \log\left(1+\frac{\delta_t}{P_t^\pi}\right).
\]
Under \cref{strat:phi},
\[
    1+\frac{\delta_t}{P_t^\pi}
    =
    \begin{cases}
        1+v_\alpha k_\alpha, & \text{with probability }\varphi,\\
        1-v_\beta k_\beta, & \text{with probability }1-\varphi.
    \end{cases}
\]
Thus the increments \(M_{t+1}^\pi-M_t^\pi\) are i.i.d.\ with mean
\[
    \mu^\pi
    =
    \varphi\log(1+v_\alpha k_\alpha)
    +(1-\varphi)\log(1-v_\beta k_\beta).
\]
By the strong law of large numbers,
\[
    \frac{M_t^\pi}{t}
    =
    \frac{M_1^\pi}{t}
    +
    \frac1t\sum_{s=1}^{t-1}
    \left(M_{s+1}^\pi-M_s^\pi\right)
    \to
    \mu^\pi
    \qquad \text{a.s.}
\]
Since \(M_1^\pi=0\), if \(\mu^\pi>0\), then
\(M_t^\pi\to+\infty\) a.s., so \(\pi\) generates persistent overpricing.
If \(\mu^\pi<0\), then \(M_t^\pi\to-\infty\) a.s., so \(\pi\) generates
persistent underpricing.

If \(\mu^\pi=0\), then \(M_t^\pi/t\to0\) a.s. If the impact is degenerate, \(v_\alpha k_\alpha=v_\beta k_\beta=0\), then \(M_t^\pi=0\) for all \(t\). Otherwise, by \cref{lem:zerodriftoscillation}, $M^\strat_t$ converges neither to \(+\infty\) nor to \(-\infty\), so \(\pi\) generates neither persistent overpricing nor persistent underpricing.
\end{proof}

\subsection{Proof of \Cref{th:zoptimumisdeltazero}}
\label{app:proofzoptimumisdeltazero}

\zoptimumisdeltazero*
\begin{proof}
The definition of $Z^\taker_t$ and $Z^\maker_t$ from \cref{eq:utilitycompetitive} shows that the sign of the utilities in \gameref{game:Z} are dictated by the sign of the price difference $\delta_t = \Et{P_{t+1} - P_t}$, while the shocks $\noise_t$ influence only the magnitude (\cref{def:priceformation}).

If the maker picks a pair $(\alpha_t, \beta_t)$ which allows for non-zero impact as per \cref{ass:concaveprice}, the taker can react by choosing a positive trade quantity $Q_t$ based on the current inventories to always achieve strictly positive utility:
\begin{itemize}
    \item If $\inv^\taker_t > \inv^\maker_t$, then by picking $Q_t > 0$, the taker guarantees $\delta_t > 0$, yielding a positive payoff for himself and a negative one for the maker.
    \item If $\inv^\taker_t < \inv^\maker_t$, the taker would pick $Q_t < 0$, guaranteeing $\delta_t < 0$ and achieving the same outcome.
\end{itemize}
Anticipating this taker's advantage, the maker's best response in a minimax sense is to restrict the feasible range of $\delta_t$ to zero by picking $\alpha_t = \beta_t = 0$. Thus, any strategy profile such that $\delta_t = 0$ for all $t$ is stable for \gameref{game:Z}.
\end{proof}

\subsection{Proof of \Cref{th:uoptimumgeneratesoverpricing}}
\label{app:proofuoptimumgeneratesoverpricing}

\uoptimumgeneratesoverpricing*

\begin{proof}
Since \(\pi\) generates persistent overpricing while \(\pi'\) does not, by \cref{th:overpricing} we have $\mu^\pi>0$ and $\mu^{\pi'}\le 0$.
Moreover,
$
    \log\nicefrac{P_t^\pi}{P_t^{\pi'}}
    =
    M_t^\pi-M_t^{\pi'}
$.
By the strong law of large numbers applied to the log-impact increments,
\[
    \frac{M_t^\pi}{t}\to \mu^\pi,
    \qquad
    \frac{M_t^{\pi'}}{t}\to \mu^{\pi'}
    \qquad \text{a.s.}
\]
Hence
\[
    \frac{1}{t}\log\frac{P_t^\pi}{P_t^{\pi'}}
    \to
    \mu^\pi-\mu^{\pi'}>0
    \qquad \text{a.s.}
\]
Therefore \(P_t^\pi>P_t^{\pi'}\) for all sufficiently large \(t\), almost
surely. Since total wealth is
\[
    W_t^\pi=C+IP_t^\pi,
    \qquad
    W_t^{\pi'}=C+IP_t^{\pi'},
\]
and \(I>0\), the same eventual ordering holds for total wealth.
\end{proof}

\subsection{Proof of \Cref{th:stoch_block_overpricing} and Stochastic Projected Gradient Ascent}
\label{app:proof_stoch_block_overpricing}

\ustochblockconv*

\begin{proof}
Fix $\gamma>0$ and $b\ge0$.  For brevity write $q_t=q_t^\gamma$, $\nu=\nu_{\gamma,b}$, $\sigma=\sigma_\gamma$, $\tau=\tau_{\gamma,b}$, and $G=G_{\gamma,b}=(b-q_0)_+$.  If $G=0$, then $q_0\ge b$ and $\tau=0$, so the claim is immediate.  Hence, in the following, assume $G>0$. Analogously, if $\sigma=0$, the argument is deterministic; thus, assume herein $\sigma>0$.

We define \emph{stopped increments} $\overline{\Delta}_{t}$ as $\overline{\Delta}_{t+1}\coloneq\Delta_{t+1}\mathbf 1\{t<\tau\}$ and the stopped martingale differences as
\[
    Z_{t+1}
    \coloneq
    \overline{\Delta}_{t+1}
    -
    \E{\overline{\Delta}_{t+1}\mid\mathcal F_t} \,,
\]
with $\E{Z_{t+1} \mid \mathcal F_t} = 0$.

Since $\{t<\tau\}\in\mathcal F_t$, the variables $Z_{t+1}$ are conditionally $\sigma$-sub-Gaussian.  Moreover, by the progress condition,
\[
    \E{\overline{\Delta}_{t+1}\mid\mathcal F_t}
    =
    \mathbf 1\{t<\tau\}\E{\Delta_{t+1}\mid\mathcal F_t}
    \ge
    \nu\mathbf 1\{t<\tau\}.
\]
On the event $\{\tau>T\}$, we have $t<\tau$ for all $t=0,\ldots,T-1$ and $q_T<b$.  Telescoping yields
\[
    \sum_{t=0}^{T-1}\overline{\Delta}_{t+1}
    =
    q_T-q_0
    <
    b-q_0
    =
    G \,,
\]
whereas
\[
    \sum_{t=0}^{T-1}\E{\overline{\Delta}_{t+1}\mid\mathcal F_t}
    \ge
    \nu T.
\]
Thus
\[
    \{\tau>T\}
    \subseteq
    \left\{
        \sum_{t=0}^{T-1} Z_{t+1}
        \le
        -(\nu T-G)
    \right\}.
\]
The standard concentration inequality for conditionally sub-Gaussian martingale differences yields, whenever $\nu T > G$,
\[
    \Pb{\tau>T}
    \le
    \exp\!\left(
        -\frac{(\nu T-G)^2}{2\sigma^2T}
    \right),
\]
with the stated convention when $\sigma=0$.

For the high-probability bound, choose
\[
    T
    \ge
    \frac{2G}{\nu}
    +
    \frac{8\sigma^2}{\nu^2}\log\frac1\delta .
\]
Then $T\ge2G/\nu$, so $\nu T-G\ge\nu T/2$, and hence
\[
    \frac{(\nu T-G)^2}{2\sigma^2T}
    \ge
    \frac{\nu^2T}{8\sigma^2}
    \ge
    \log\frac1\delta .
\]
Therefore $\Pb{\tau>T}\le\delta$.  Since the same bound tends to zero as $T\to\infty$, we obtain $\tau<\infty$ almost surely.

Taking $b=0$, we have
\[
    \tau_{\gamma,0}
    =
    \inf\{t\ge0:q_t^\gamma\ge0\}
    =
    \inf\{t\ge0:\mu^\strat_t\ge\gamma\},
\]
so the scheme reaches overpricing strength $\gamma$ in finite time almost surely, and $G_{\gamma,0}=(-q_0^\gamma)_+=(\gamma-\mu^\strat_0)_+$.

It remains to prove the optional buffered persistence statement.  Assume $b>0$ and condition on $\mathcal F_{\tau_{\gamma,b}}$ on the event $\{\tau_{\gamma,b}<\infty\}$.  At time $\tau_{\gamma,b}$, $q_{\tau_{\gamma,b}}^\gamma\ge b$.  Let $\chi\coloneq\inf\{h\ge0:q_{\tau_{\gamma,b}+h}^\gamma<0\}$ be the first exit time from the overpricing region.  On the event $\{\chi\le H\}$, before exit the process stays in $\{q^\gamma\ge0\}$, where the additional stability assumption gives nonnegative conditional drift.  Therefore, exit by time $H$ requires the centered martingale noise accumulated over at most $H$ steps to fall below $-b$.  By the maximal inequality for conditionally sub-Gaussian martingales,
\[
    \Pb{\chi\le H\mid\mathcal F_{\tau_{\gamma,b}}}
    \le
    \exp\!\left(
        -\frac{b^2}{2\sigma_\gamma^2H}
    \right).
\]
Equivalently,
\[
    \Pb{
        \min_{0\le h\le H}q_{\tau_{\gamma,b}+h}^\gamma\ge0
        \,\middle|\,
        \mathcal F_{\tau_{\gamma,b}}
    }
    \ge
    1-\exp\!\left(-\frac{b^2}{2\sigma_\gamma^2H}\right).
\]
This concludes the proof.
\end{proof}

\subsubsection{Stochastic Projected Gradient Ascent}
\label{app:psga}

We now verify the stochastic block-coordinate argument for projected
stochastic gradient ascent.  For this concrete algorithm, it is useful to work
with the product-space certificate $s_t^\gamma=x_t-r_\gamma(y_t)$, whose
nonnegativity is equivalent to overpricing strength $\gamma$.
We will rely on the following lemma.

\begin{lemma}[Equivalent criteria for overpricing strength]\label{lemma:overpricing_equivalent_criteria}
    Fix $\phi \in (0,1)$ and $\gamma \in (0, \infty)$, define:
    \begin{equation}
        r_\gamma(y)
    \coloneq
        e^{\frac{\gamma}{\phi}} (1-y)^{-\frac{1-\phi}{\phi}} - 1
    \quad \text{and} \quad
        g_\gamma(x)
    \coloneq
        1 - e^{\frac{\gamma}{1-\phi}} (1+x)^{-\frac{\phi}{1-\phi}} \,.
    \end{equation}
    For any choice of parameters $(\ka, \kb, \va, \vb)$ in the feasible region and for any choice $\gamma > 0$, the following are equivalent:
    \[
        \mu^\strat \ge \gamma
    \iff
          x \ge r_\gamma(y)
    \iff
          y \leq g_\gamma(x) \,,
    \]
    where $x = \ka \va$, $y = \kb \vb$ and $\mu^\strat$ is defined in \cref{eq:mu_overpricing}.
\end{lemma}
\begin{proof}
Recall the definition $\mu^\strat = \phi \log(1 + \va \ka) + (1 - \phi) \log(1 - \vb \kb)$ from \cref{eq:mu_overpricing}, thus 
\begin{align*}
    \phi \log(1 + \va \ka) + (1 - \phi) \log(1 - \vb \kb) \ge \gamma
    \iff
     \va \ka \ge e^{\frac{\gamma}{\phi}} (1 - \vb \kb)^{- \frac{(1 - \phi)}{\phi}} -1  \,, 
\end{align*}
where we used $\phi \in (0,1)$. Denoting $x\coloneq \va\ka$ and $y\coloneq \vb\kb$ yields the equivalent criteria $x \ge r_\gamma(y)$ with $r_\gamma(y) \coloneq e^{\frac{\gamma}{\phi}}(1-y)^{-\frac{1-\phi}{\phi} }-1$. 
The second equivalence follows by noticing $g_\gamma \equiv r_\gamma^{-1}$.
\end{proof}

We notice that as $y$ decreases, $r_\gamma(y)$ also decreases, making the condition easier to satisfy. Since the increase of $x$ affects the criteria linearly, when $y$ is closer to $1$, the $y$ updates (the $\beta$-block) are more effective toward satisfying the criterion and, as $y$ decreases, the $x$ updates become more effective.

Thus, we have that, for
$r_\gamma(y)=e^{\gamma/\phi}(1-y)^{-(1-\phi)/\phi}-1$, the condition
$\mu^\strat_t\ge\gamma$ is equivalent to $x_t\ge r_\gamma(y_t)$.  Hence,
define the product-space slack $s_t^\gamma\coloneq x_t-r_\gamma(y_t)$.

For $\epsilon_\pi\ge0$, let
\[
    \Pi_{\alpha,\epsilon_\pi}(\va,z)
    \coloneq
    \Pi_{[0,f_\alpha(\va)-\epsilon_\pi]}(z),
    \qquad
    \Pi_{\beta,\epsilon_\pi}(\vb,z)
    \coloneq
    \Pi_{[0,f_\beta(\vb)-\epsilon_\pi]}(z),
\]
whenever the intervals are nonempty.  

We now state the \emph{projected stochastic gradient ascent algorithm}.

\begin{algorithmdef}[Projected stochastic gradient ascent]
\label{def:psga}
Fix stepsizes $\eta_{\va},\eta_{\ka},\eta_{\vb},\eta_{\kb}>0$ and an initial
feasible profile.  At time $t$, independently of $\mathcal F_t$, select the
$\alpha$-block with probability $\phi$ and the $\beta$-block with probability
$1-\phi$.  If the $\alpha$-block is selected, draw noise
$(\xi_{\va,t},\xi_{\ka,t})$ and set
\[
    \va^{t+1}=\va^t+\eta_{\va}(\ka^t+\xi_{\va,t}),
    \qquad
    \ka^{t+1}=
    \Pi_{\alpha,\epsilon_\pi}\!\left(
        \va^{t+1},
        \ka^t+\eta_{\ka}(\va^t+\xi_{\ka,t})
    \right).
\]
If the $\beta$-block is selected, draw noise
$(\xi_{\vb,t},\xi_{\kb,t})$ and set
\[
    \vb^{t+1}=\vb^t-\eta_{\vb}(\kb^t+\xi_{\vb,t}),
    \qquad
    \kb^{t+1}=
    \Pi_{\beta,\epsilon_\pi}\!\left(
        \vb^{t+1},
        \kb^t-\eta_{\kb}(\vb^t+\xi_{\kb,t})
    \right).
\]
All iterates are assumed feasible almost surely.
\end{algorithmdef}

Let $\overline\strat^{t+1}$ denote the one-step update of \cref{def:psga} with the
same block choice as $\strat^{t+1}$ but with all gradient noise set to zero.
Let $\overline s_{t+1}^\gamma$ be the corresponding noiseless product-space slack,
and define
\[
    d_{t+1}^\gamma\coloneq s_{t+1}^\gamma-s_t^\gamma,
    \qquad
    \overline d_{t+1}^\gamma\coloneq \overline s_{t+1}^\gamma-s_t^\gamma .
\]

\begin{assumption}[Projected stochastic-gradient progress]
\label{ass:psga_product_progress}
Fix $\gamma>0$ and $b\ge0$.  There exist constants
$m_{\gamma,b}^{\pi}>0$, $\rho_{\gamma,b}^{\pi}\in[0,m_{\gamma,b}^{\pi})$, and
$\sigma_{\gamma}^{\pi}<\infty$ such that, for all $t$,
\[
    \E{\overline d_{t+1}^\gamma\mid\mathcal F_t}
    \ge
    m_{\gamma,b}^{\pi}\mathbf 1\{s_t^\gamma<b\},
    \qquad
    \E{d_{t+1}^\gamma-\overline d_{t+1}^\gamma\mid\mathcal F_t}
    \ge
    -\rho_{\gamma,b}^{\pi}\mathbf 1\{s_t^\gamma<b\},
\]
and $d_{t+1}^\gamma-\E{d_{t+1}^\gamma\mid\mathcal F_t}$ is conditionally
$\sigma_\gamma^\pi$-sub-Gaussian.
\end{assumption}

\begin{restatable}[Projected stochastic gradient ascent reaches overpricing]{corollary}{upsgaoverpricing}
\label{cor:psga_overpricing}
Fix $\phi\in(0,1)$, $\gamma>0$, and $b\ge0$.  Let $\strat^t$ evolve according
to \cref{def:psga}, and suppose \cref{ass:psga_product_progress} holds.  Set
$\nu_{\gamma,b}^{\pi}\coloneq m_{\gamma,b}^{\pi}-\rho_{\gamma,b}^{\pi}>0$,
$\tau_{\gamma,b}^{\pi}\coloneq\inf\{t\ge0:s_t^\gamma\ge b\}$, and
$G_{\gamma,b}^{\pi}\coloneq(b-s_0^\gamma)_+$.  Then, for every $T\ge1$ with
$\nu_{\gamma,b}^{\pi}T>G_{\gamma,b}^{\pi}$,
\[
    \Pb{\tau_{\gamma,b}^{\pi}>T}
    \le
    \exp\!\left(
        -\frac{(\nu_{\gamma,b}^{\pi}T-G_{\gamma,b}^{\pi})^2}
        {2(\sigma_\gamma^\pi)^2T}
    \right).
\]
Consequently, $\tau_{\gamma,b}^{\pi}<\infty$ almost surely, and for every
$\delta\in(0,1)$, with probability at least $1-\delta$,
\[
    \tau_{\gamma,b}^{\pi}
    \le
    \left\lceil
        \frac{2G_{\gamma,b}^{\pi}}{\nu_{\gamma,b}^{\pi}}
        +
        \frac{8(\sigma_\gamma^\pi)^2}{(\nu_{\gamma,b}^{\pi})^2}
        \log\frac1\delta
    \right\rceil .
\]
In particular, taking $b=0$, projected stochastic gradient ascent reaches
overpricing strength at least $\gamma$ in finite time almost surely, with
$G_{\gamma,0}^{\pi}=(r_\gamma(y_0)-x_0)_+$.
\end{restatable}

\begin{proof}
Write $x_t=\va^t\ka^t$, $y_t=\vb^t\kb^t$, and
$s_t^\gamma=x_t-r_\gamma(y_t)$.  By
\cref{lemma:overpricing_equivalent_criteria},
\[
    s_t^\gamma\ge0
    \iff
    x_t\ge r_\gamma(y_t)
    \iff
    \mu^\strat_t\ge\gamma .
\]
Thus hitting the set $\{s^\gamma\ge0\}$ is equivalent to reaching
overpricing strength at least $\gamma$.

The update in \cref{def:psga} is a stochastic randomized block-coordinate
scheme: at each iteration only the selected block is updated, the other block
is unchanged, and the strict projection enforces feasibility.  Projection is
already included in both the noisy update $s_{t+1}^\gamma$ and the noiseless
comparison update $\overline s_{t+1}^\gamma$.

Let
\[
    d_{t+1}^\gamma=s_{t+1}^\gamma-s_t^\gamma,
    \qquad
    \overline d_{t+1}^\gamma=\overline s_{t+1}^\gamma-s_t^\gamma .
\]
By \cref{ass:psga_product_progress},
\[
    \E{d_{t+1}^\gamma\mid\mathcal F_t}
    =
    \E{\overline d_{t+1}^\gamma\mid\mathcal F_t}
    +
    \E{d_{t+1}^\gamma-\overline d_{t+1}^\gamma\mid\mathcal F_t}
    \ge
    \nu_{\gamma,b}^{\pi}\mathbf 1\{s_t^\gamma<b\},
\]
where $\nu_{\gamma,b}^{\pi}=m_{\gamma,b}^{\pi}-\rho_{\gamma,b}^{\pi}>0$.
Moreover, the centered increment
$d_{t+1}^\gamma-\E{d_{t+1}^\gamma\mid\mathcal F_t}$ is conditionally
$\sigma_\gamma^\pi$-sub-Gaussian.

Define the buffered hitting time
\[
    \tau_{\gamma,b}^{\pi}
    \coloneq
    \inf\{t\ge0:s_t^\gamma\ge b\},
    \qquad
    G_{\gamma,b}^{\pi}
    \coloneq
    (b-s_0^\gamma)_+ .
\]
If $G_{\gamma,b}^{\pi}=0$, then $\tau_{\gamma,b}^{\pi}=0$.  Otherwise define
the stopped increments
\[
    \widetilde d_{t+1}^\gamma
    \coloneq
    d_{t+1}^\gamma\mathbf 1\{t<\tau_{\gamma,b}^{\pi}\}
\]
and the stopped martingale differences
\[
    Z_{t+1}
    \coloneq
    \widetilde d_{t+1}^\gamma
    -
    \E{\overline d_{t+1}^\gamma\mid\mathcal F_t}.
\]
Since $\{t<\tau_{\gamma,b}^{\pi}\}\in\mathcal F_t$, the variables
$Z_{t+1}$ are conditionally $\sigma_\gamma^\pi$-sub-Gaussian.  Also,
\[
    \E{\widetilde d_{t+1}^\gamma\mid\mathcal F_t}
    =
    \mathbf 1\{t<\tau_{\gamma,b}^{\pi}\}
    \E{d_{t+1}^\gamma\mid\mathcal F_t}
    \ge
    \nu_{\gamma,b}^{\pi}\mathbf 1\{t<\tau_{\gamma,b}^{\pi}\}.
\]

On the event $\{\tau_{\gamma,b}^{\pi}>T\}$, we have
$t<\tau_{\gamma,b}^{\pi}$ for all $t=0,\ldots,T-1$ and
$s_T^\gamma<b$.  Therefore,
\[
    \sum_{t=0}^{T-1}\widetilde d_{t+1}^\gamma
    =
    s_T^\gamma-s_0^\gamma
    <
    b-s_0^\gamma
    =
    G_{\gamma,b}^{\pi},
\]
whereas
\[
    \sum_{t=0}^{T-1}
    \E{\widetilde d_{t+1}^\gamma\mid\mathcal F_t}
    \ge
    \nu_{\gamma,b}^{\pi}T.
\]
Hence
\[
    \{\tau_{\gamma,b}^{\pi}>T\}
    \subseteq
    \left\{
        \sum_{t=0}^{T-1}Z_{t+1}
        \le
        -(\nu_{\gamma,b}^{\pi}T-G_{\gamma,b}^{\pi})
    \right\}.
\]
The concentration inequality for conditionally sub-Gaussian martingale
differences gives, whenever
$\nu_{\gamma,b}^{\pi}T>G_{\gamma,b}^{\pi}$,
\[
    \Pb{\tau_{\gamma,b}^{\pi}>T}
    \le
    \exp\!\left(
        -\frac{(\nu_{\gamma,b}^{\pi}T-G_{\gamma,b}^{\pi})^2}
        {2(\sigma_\gamma^\pi)^2T}
    \right).
\]
The stated high-probability bound follows by choosing
\[
    T
    \ge
    \frac{2G_{\gamma,b}^{\pi}}{\nu_{\gamma,b}^{\pi}}
    +
    \frac{8(\sigma_\gamma^\pi)^2}{(\nu_{\gamma,b}^{\pi})^2}
    \log\frac1\delta .
\]
Indeed, this implies
$\nu_{\gamma,b}^{\pi}T-G_{\gamma,b}^{\pi}\ge\nu_{\gamma,b}^{\pi}T/2$
and therefore the exponent is at least $\log(1/\delta)$.  Letting
$T\to\infty$ gives $\tau_{\gamma,b}^{\pi}<\infty$ almost surely.

Finally, when $b=0$,
\[
    \tau_{\gamma,0}^{\pi}
    =
    \inf\{t\ge0:s_t^\gamma\ge0\}
    =
    \inf\{t\ge0:\mu^\strat_t\ge\gamma\},
\]
by \cref{lemma:overpricing_equivalent_criteria}.  Moreover,
\[
    G_{\gamma,0}^{\pi}
    =
    (-s_0^\gamma)_+
    =
    (r_\gamma(y_0)-x_0)_+ .
    \qedhere
\]
\end{proof}

\subsection{Proof of \Cref{lem:farsightedobjective}}
\label{app:prooffarsightedobjective}

The proof of \cref{lem:farsightedobjective} relies on the following auxiliary lemma which let's us control the expected price ratio in the price-inflating regime under \cref{ass:boundednoise}.
\begin{restatable}{lemma}{expectedpriceratio}\label{lem:expectedpriceratio}
    Assume that \cref{ass:boundednoise} holds and, for all $t$, $\E{\log Y_{t+1}} > 0$, where $Y_{t+1} \coloneq \frac{P_{t+1}}{P_t}$ and $(Y_s)_{s}$ are i.i.d., then there exists $a \in (0, d \wedge 1]$ such that, for all $t$, $\E{ Y_{t+1}^{-a} } < 1.$
\end{restatable}

\begin{proof}
Let $X \coloneq \log Y$ and define $g(a) \coloneq \E{Y^{-a}} = \E{e^{-aX}}$. By \cref{ass:boundednoise} and the boundedness away from zero and infinity of \(1+\delta_t/P_t\) under \cref{strat:phi}, \(g(a)<\infty\) for all \(a\in[-d,d]\).
We first note that $\E{|X|} < \infty$. Indeed, for any $b \in (0,d]$ and any $x > 0$,
\[
|\log x| \le \frac{x^{b}+x^{-b}}{b},
\]
hence $\E{|X|} \le \frac{1}{b} (\E{Y^{b}} + \E{Y^{-b}}) < \infty$.
Fix $a_0\in(0,d/2]$. For $a\in(0,a_0]$,
\[
    \frac{g(a)-g(0)}{a}=\E{\frac{e^{-aX}-1}{a}}.
\]
To analyze the integrand, fix an outcome of $X$, i.e., set $x \coloneq X$ and consider $f(u)=e^{-ux}$ on $[0,a]$.
By the mean value theorem, there exists $c\in(0,a)$ such that
\[
\frac{e^{-ax}-1}{a}=\frac{f(a)-f(0)}{a-0}=f'(c)=-x e^{-cx}.
\]
Writing $c=\theta a$ with $\theta\in(0,1)$ gives the point-wise identity
\[
\frac{e^{-aX}-1}{a}=-X\,e^{-\theta aX}.
\]

We now verify the conditions of the dominated convergence theorem to justify exchanging limit and expectation.
First, for each fixed outcome $x$, it holds that
\[
    \frac{e^{-ax}-1}{a}\to -x \qquad \text{as} \quad a \to 0.
\]
Second, we bound uniformly in $a\in(0,a_0]$. Since $e^{-\theta aX}=Y^{-\theta a}$, for any $\theta\in(0,1)$,
$
Y^{-\theta a}\le Y^{a_0}+Y^{-a_0},
$
hence
\[
    \labs{\frac{e^{-aX}-1}{a}}
\le
    |X| (Y^{a_0}+Y^{-a_0}).
\]
Moreover, using $|X|\le \frac{1}{a_0}(Y^{a_0}+Y^{-a_0})$, we get
\[
    |X| (Y^{a_0}+Y^{-a_0})
\le
    \frac{1}{a_0}(Y^{a_0}+Y^{-a_0})^2
\le
    \frac{2}{a_0}(Y^{2a_0}+Y^{-2a_0})
\eqcolon
    Z.
\]
By \cref{ass:boundednoise} and $2a_0\le d$, we have $\E{Z} < \infty$. Therefore dominated convergence applies and
\[
    g'(0)
=
    \lim_{a \to 0}\E{\frac{e^{-aX}-1}{a}}
=
    \E{\lim_{a \to 0}\frac{e^{-aX}-1}{a}}
=
    \E{-X}
=
    -\E{\log Y}
<
    0.
\]
Since $g(0)=1$ and $g$ is continuous at $0$ with negative right-derivative, there exists
$a\in(0,\min\{a_0,1\}]$ such that $g(a) = \E{Y^{-a}} < 1$.
\end{proof}

We are now ready to prove \cref{lem:farsightedobjective}.

\farsightedobjective*
\begin{proof}
Start by considering the following decomposition
\[
    \log \wealth^p_t
=
    \log\lrb{ P_t \inv^p_t + \cash^p_t }
=
    \log\lrb{ P_t \inv^p_t \lrb{1 + \frac{\cash^p_t}{P_t \inv^p_t} }}
=
    \log P_t + \log \inv^p_t + \log \lrb{1 + \frac{\cash^p_t}{P_t \inv^p_t} } \,.
\]
Which we can use to write the farsighted objective as
\begin{align*}
    J^p_\strat
&=
    \lim_{T \to \infty} \frac{1}{T} \E{
        \log P_T + \log \inv^p_T + \log \lrb{1 + \frac{\cash^p_T}{P_T \inv^p_T} }
    } - \frac{1}{T} \log \wealth^p_1
\\ &=
    \lim_{T \to \infty}
        \frac{1}{T} \underbrace{\E{ \log P_T }}_{(\mathrm{I})}
        + \frac{1}{T} \underbrace{\E{ \log \inv^p_T }}_{(\mathrm{II})}
        + \frac{1}{T} \underbrace{\E{ \log \lrb{1 + \frac{\cash^p_T}{P_T \inv^p_T} } }}_{(\mathrm{III})} \,.
\end{align*}
We make considerations on each term individually:
\begin{enumerate}
    \item[(I)] By definition of price (see \cref{th:overpricing}), we can simplify term $(\mathrm{I})$ as
    \begin{align*}
        (\mathrm{I})
    &=
        \E{ \log P_T }
    =
        \E{ \log P_T - \log P_1 } + \log P_1
    \\ &=
        \sum_{t=1}^{T-1} \E{ \log \frac{P_{t+1}}{P_t} } + \log P_1
    =
        \sum_{t=1}^{T-1} \E{ \log \eta_t + \log \noise_{t+1} } + \log P_1
    \\ &=
        (T - 1)(\mu^\strat + \mu^\noise) + \log P_1
    \end{align*}
    Thus, $\frac{1}{T}(\mathrm{I}) \to \mu^\strat + \mu^\noise$.
    \item[(II)] Since $i \le \inv^p_T \le I$ a.s. by feasibility, where $i > 0$ by assumption, we have $\log i \le \log \inv^p_T \le \log I$ a.s., hence $(\mathrm{II}) = \E{\log \inv^p_T}$ is uniformly bounded in $T$ and thus $\frac{1}{T}(\mathrm{II}) \to 0$.
    \item[(III)] By feasibility, we can show that the term $(\mathrm{III})$ is asymptotically dominated by the price
    \[
        \E{ \log \lrb{1 + \frac{1}{P_T} \cdot \frac{c}{I} } }
    \le
        (\mathrm{III})
    \le
        \E{ \log \lrb{1 + \frac{1}{P_T} \cdot \frac{C}{i} } }
    \]
\end{enumerate}

Following the structure of \cref{th:overpricing}, we study the value of the farsighted objective under the different regimes of the price process $P_t$:
\begin{itemize}
    \item If $P_t \to \infty$ a.s., making use of \cref{lem:expectedpriceratio}, call $Y_{t+1} = \nicefrac{P_{t+1}}{P_t}$ i.i.d.\ and $Y = Y_t$ for all $t$, then
    \begin{align*}
        (\mathrm{III})
    &\le
        \E{ \log \lrb{1 + \frac{1}{P_T} \cdot \frac{C}{i} } }
    \\ &\le \tag{$\log(1 + x) \le \frac{x^a}{a}$ for all $a \in (0, 1]$ and $x > 0$}
        \frac{1}{a} \lrb{ \frac{C}{i} }^a \E{ P_T^{-a} }
    \\ &=
        \frac{1}{a} \lrb{ \frac{C}{i P_1} }^a \E{ \prod_{t=1}^{T-1} \lrb{\frac{P_{t+1}}{P_t}}^{-a} }
    \\ &=
        \frac{1}{a} \lrb{ \frac{C}{i P_1} }^a \E{ \prod_{t=1}^{T-1} Y_{t+1}^{-a} }
    \\ &= \tag{$Y_t$ i.i.d.}
        \frac{1}{a} \lrb{ \frac{C}{i P_1} }^a \E{Y^{-a}}^{T-1}
    \end{align*}
    which decays exponentially fast in $T$ by \cref{lem:expectedpriceratio} since $\E{Y^{-a}} < 1$. In particular, $(\mathrm{III}) = O(\E{Y^{-a}}^{T})$ and therefore $\frac{1}{T}(\mathrm{III}) \to 0$. 
    Thus $J^p_\strat = \mu^\strat + \mu^\noise$.

    \item If $P_t \to 0$ a.s., since \(\log(1+x)=\log x+\log(1+1/x)\), we have
    \[
    \begin{aligned}
        (\mathrm{III})
        &=
        \E{\log\lrb{1+\frac{C_T^p}{P_T I_T^p}}} \\
        &=
        -\E{\log P_T}
        +
        \E{\log C_T^p-\log I_T^p}
        +
        \E{\log\lrb{1+\frac{P_T I_T^p}{C_T^p}}}.
    \end{aligned}
    \]
    By feasibility, \(c\le C_T^p\le C\) and \(i\le I_T^p\le I\), hence
    \[
        \log\frac{c}{I}
        \le
        \E{\log C_T^p-\log I_T^p}
        \le
        \log\frac{C}{i},
    \]
    and
    \[
        0
        \le
        \E{\log\lrb{1+\frac{P_T I_T^p}{C_T^p}}}
        \le
        \E{\log\lrb{1+\frac{I}{c}P_T}}.
    \]
    Let \(Y_{t+1}:=P_{t+1}/P_t\). In the regime \(P_T\to0\) a.s., we have
    \(\E{\log Y_{t+1}}<0\). By \cref{ass:boundednoise}, the same argument as in
    \cref{lem:expectedpriceratio} gives some \(a\in(0,d\wedge 1]\) such that
    \(\E{Y_{t+1}^a}<1\). Therefore
    \[
        \E{\log\lrb{1+\frac{I}{c}P_T}}
        \le
        \frac{1}{a}\lrb{\frac{I}{c}}^a \E{P_T^a}
        =
        \frac{1}{a}\lrb{\frac{I P_1}{c}}^a
        \E{Y_{t+1}^a}^{T-1}
        \to 0 .
    \]
    Thus
    \[
        \frac{1}{T}(\mathrm{III})
        =
        -\frac{1}{T}(\mathrm{I})+o(1).
    \]
    Hence \((\mathrm{I})\) and \((\mathrm{III})\) cancel asymptotically, while
    \((\mathrm{II})/T\to0\), so \(J^p_\strat=0\).
\end{itemize}
By the strong law of large numbers applied to \(\log(P_{t+1}/P_t)\), if \(\mu^\pi+\mu^\varepsilon>0\) then \(P_t\to\infty\) almost surely, while if \(\mu^\pi+\mu^\varepsilon<0\) then \(P_t\to0\) almost surely.

The only remaining case is $\mu^\pi + \mu^\varepsilon = 0$, in which \((\mathrm{I})/T\to 0\) and it remains to show that \((\mathrm{III})/T\to0\). By feasibility,
\[
    0\le (\mathrm{III})
    \le
    \E{\log\lrb{1+\frac{C}{iP_T}}}.
\]
Moreover,
\[
    \log\lrb{1+\frac{C}{iP_T}}
    =
    \log\lrb{1+\frac{C}{i}e^{-\log P_T}}
    \le
    \log\lrb{1+\frac{C}{i}}+(\log P_T)^- ,
\]
where \(x^-:=\max\{-x,0\}\). Since
\[
    \log P_T
    =
    \log P_1+\sum_{t=1}^{T-1}\log\frac{P_{t+1}}{P_t},
    \qquad
    \E{\log\frac{P_{t+1}}{P_t}}=0,
\]
and
$
    \nicefrac{P_{t+1}}{P_t}
    =
    (1+\nicefrac{\delta_t}{P_t})\varepsilon_{t+1}
$, the increments \(\log(P_{t+1}/P_t)\) have finite second moment by
\cref{ass:boundednoise} and the boundedness of \(1+\delta_t/P_t\) under
\cref{strat:phi}. Hence
\[
    \E{(\log P_T)^-}
    \le
    \E{|\log P_T|}
    =
    O(\sqrt{T}).
\]
Thus \((\mathrm{III})/T\to0\). Since also \((\mathrm{I})/T\to0\) and
\((\mathrm{II})/T\to0\), we get \(J^p_\strat=0\).

In conclusion, $J^p_\pi=\max\{\mu^\pi+\mu^\varepsilon,0\}$.
\end{proof}

\subsection{Proof of \Cref{th:myopicoptimizesfarsighted}}
\label{app:proofmyopicoptimizesfarsighted}

\myopicoptimizesfarsighted*

\begin{proof}
By \cref{th:stoch_block_overpricing}, the iterates reach a profile \(\pi\) such that \(\mu^\pi\ge \mu^\star-\epsilon\) in finite time almost surely. By \cref{lem:farsightedobjective},
\[
    J^p_\pi
    =
    \max\{\mu^\pi+\mu^\varepsilon,0\}
    \ge
    \max\{\mu^\star-\epsilon+\mu^\varepsilon,0\}
    \ge
    J^\star-\epsilon .
    \qedhere
\]
\end{proof}

\end{document}